\documentclass[zpreprint,zbstpl]{./zeus_paper}
\usepackage[english]{babel}
\chardef\usc=95
\chardef\til=126
\catcode`\@=11 
\DeclareRobustCommand\xdotspace{\futurelet\@let@token\@xdotspace}
\def\@xdotspace{%
  \ifx\@let@token.\else
  \ifx\@let@token\bgroup.\else
  \ifx\@let@token\egroup.\else
  \ifx\@let@token\/.\else
  \ifx\@let@token\ .\else
  \ifx\@let@token~.\else
  \ifx\@let@token!.\else
  \ifx\@let@token,.\else
  \ifx\@let@token:.\else
  \ifx\@let@token;.\else
  \ifx\@let@token?.\else
  \ifx\@let@token/.\else
  \ifx\@let@token'.\else
  \ifx\@let@token).\else
  \ifx\@let@token-.\else
  \ifx\@let@token\@xobeysp.\else
  \ifx\@let@token\space.\else
  \ifx\@let@token\@sptoken.\else
   .\space
   \fi\fi\fi\fi\fi\fi\fi\fi\fi\fi\fi\fi\fi\fi\fi\fi\fi\fi}
\catcode`\@=12 

\newcommand{\stru}[2]{%
   \relax\ifmmode\hbox{\vrule height#1 depth#2 width0pt}%
   \else\vrule height#1 depth#2 width0pt\fi}

\newcommand{\Ronum}[1]{\uppercase\expandafter{\romannumeral#1}}
\newcommand{\ronum}[1]{\expandafter{\romannumeral#1}}
\DeclareRobustCommand{\LaTeXZ}{%
  \LaTeX\kern-.05em4\kern-.1em
  {\raisebox{-0.2ex}{$\scriptstyle\text{ZEUS}$}}\xspace}

\DeclareMathAlphabet{\mathbf}{OT1}{cmr}{bx}{sl}
\newcommand{\eVdist}{\kern-0.06667em}

\newcommand{\Gev}{{\text{Ge}\eVdist\text{V\/}}}

\newcommand{\gev}{{\,\text{Ge}\eVdist\text{V\/}}}

\newcommand{\met}{\,\text{m}}
\newcommand{\nm}{\,\text{nm}}

\newcommand{\mm}{\,\text{mm}}
\newcommand{\cm}{\,\text{cm}}
\newcommand{\km}{\,\text{km}}

\newcommand{\Hz}{\,\text{Hz}}
\newcommand{\kHz}{\,\text{kHz}}
\newcommand{\MHz}{\,\text{MHz}}
\newcommand{\GHz}{\,\text{GHz}}

\newcommand{\rad}{\,\text{rad}}

\newcommand{\mrad}{\,\text{mrad}}

\newcommand{\slashfrac}[2]{%
  \raisebox{0.5ex}{\ensuremath #1}\kern-0.12em/\kern-0.08em
  \raisebox{-.8ex}{\ensuremath #2}}

\newcommand{\sqr}[3]{%
    {\vcenter{\hrule height.#3ex\hbox{\vrule width.#2ex height#1ex
     \kern#1ex\vrule width.#3ex}\hrule height.#2ex}}}

\catcode`\@=11 
\newcommand{\parenbar}{\mathpalette\p@renb@r}
\def\p@renb@r#1#2{\vbox{%
  \ifx#1\scriptscriptstyle \dimen@.7em\dimen@ii.2em\else
  \ifx#1\scriptstyle \dimen@.8em\dimen@ii.25em\else
  \dimen@1em\dimen@ii.4em\fi\fi \offinterlineskip
  \ialign{\hfill##\hfill\cr
    \vbox{\hrule width\dimen@ii}\cr
    \noalign{\vskip-.3ex}%
    \hbox to\dimen@{$\mathchar300\hfil\mathchar301$}\cr
    \noalign{\vskip-.3ex}%
    $#1#2$\cr}}}
\catcode`\@=12 

\newcommand{\IP}{{\rm I$\kern-0.01667em$P}\xspace}
\newcommand{\JB}{{\rm JB}}

\newcommand{\jet}{{\rm jet}}

\mathchardef\qsm=63
\mathchardef\pls=43
\mathchardef\mns=512
\mathchardef\plm=518
\mathchardef\eql=61
\mathchardef\smallleft=300
\mathchardef\smallright=301
\mathchardef\les=316
\mathchardef\gre=318
\mathchardef\leq=532
\mathchardef\grq=533

\catcode`\@=11 
\newcounter{pict@width}
\newcounter{pict@height}
\newlength{\pict@scale}
\newcommand{\psfigadd}[4]{%
\setcounter{pict@width}{1*\ratio{#2+\pict@scale/2}{\pict@scale}}
\setcounter{pict@height}{1*\ratio{#3+\pict@scale/2}{\pict@scale}}
\setlength{\unitlength}{\pict@scale}
\hbox to #2{\hspace{-\fill}\begin{picture}(\thepict@width,\thepict@height)
\put(0,0){\psfig{figure=#1,width=#2,height=#3,clip=}}
\SetScale{0.283466457}
\SetWidth{1.763889}
{#4}
\end{picture}}
}
\newcounter{pict@widthfst}
\newcounter{pict@widthscd}
\newcounter{pict@widthtot}
\newcommand{\psfigaddtwo}[7]{%
\setcounter{pict@widthfst}{1*\ratio{#2+\pict@scale/2}{\pict@scale}}
\setcounter{pict@widthscd}{1*\ratio{#2+#4+\pict@scale/2}{\pict@scale}}
\setcounter{pict@widthtot}{1*\ratio{#2+#4+#6+\pict@scale/2}{\pict@scale}}
\setcounter{pict@height}{1*\ratio{#3+\pict@scale/2}{\pict@scale}}
\setlength{\unitlength}{\pict@scale}
\hbox{\hspace{-\fill}\begin{picture}(\thepict@widthtot,\thepict@height)
\put(0,0){\psfig{figure=#1,width=#2,height=#3,clip=}}
\put(\thepict@widthscd,0){\psfig{figure=#5,width=#6,height=#3,clip=}}
\SetScale{0.283466457}
\SetWidth{1.763889}
{#7}
\end{picture}}
}
\newcommand{\psfigror}[4]{%
\setcounter{pict@width}{1*\ratio{#2+\pict@scale/2}{\pict@scale}}
\setcounter{pict@height}{1*\ratio{#3+\pict@scale/2}{\pict@scale}}
\setlength{\unitlength}{\pict@scale}
\hbox{\begin{picture}(\thepict@width,\thepict@height)
\put(0,\thepict@height){\psfig{figure=#1,width=#3,height=#2,clip=,angle=270}}
\SetScale{0.283466457}
\SetWidth{1.763889}
{#4}
\end{picture}}
}
\newcommand{\psfigrol}[4]{%
\setcounter{pict@width}{1*\ratio{#2+\pict@scale/2}{\pict@scale}}
\setcounter{pict@height}{1*\ratio{#3+\pict@scale/2}{\pict@scale}}
\setlength{\unitlength}{\pict@scale}
\hbox{\begin{picture}(\thepict@width,\thepict@height)
\put(0,0){\psfig{figure=#1,width=#3,height=#2,clip=,angle=90}}
\SetScale{0.283466457}
\SetWidth{1.763889}
{#4}
\end{picture}}
}
\catcode`\@=12 
\newlength\listtextwidth

\catcode`\@=11 
\newlength{\@tabfninsert}
\newlength{\@tabfnwidth}
\newcommand{\tabfootnote}[2]{%
  \setlength{\@tabfninsert}{0.8em}
  \setlength{\@tabfnwidth}{\textwidth}
  \addtolength{\@tabfnwidth}{-\@tabfninsert}
  \addtolength{\@tabfnwidth}{-0.4em}
  \noindent\makebox[\@tabfninsert][r]{\footnotesize$^{#1}$\hfil}\hfill%
  \parbox[t]{\@tabfnwidth}{\footnotesize #2\hfill}}
\catcode`\@=12 

\newcommand{\ZcoosysAA}{%
The ZEUS coordinate system is a right-handed Cartesian system, with the $Z$
axis pointing in the proton beam direction, referred to as the ``forward
direction'', and the $X$ axis pointing  towards the center of HERA.
The coordinate origin is at the nominal interaction point.\xspace}
\newcommand{\ZcoosysAE}{%
The ZEUS coordinate system is a right-handed Cartesian system, with the $Z$
axis pointing in the proton beam direction, referred to as the ``forward
direction'', and the $X$ axis pointing  towards the centre of HERA.
The coordinate origin is at the nominal interaction point.\xspace}
\newcommand{\ZcoosysBA}{%
The ZEUS coordinate system is a right-handed Cartesian system, with the $Z$ 
axis pointing in the nominal proton beam direction, referred to as the ``forward
direction'', and the $X$ axis pointing  towards the center of HERA.
The coordinate origin is at the centre of the CTD.\xspace}
\newcommand{\ZcoosysBE}{%
The ZEUS coordinate system is a right-handed Cartesian system, with the $Z$ 
axis pointing in the nominal proton beam direction, referred to as the ``forward
direction'', and the $X$ axis pointing  towards the centre of HERA.
The coordinate origin is at the centre of the CTD.\xspace}
\newcommand{\ZcoosysCE}{%
The ZEUS coordinate system is a right-handed Cartesian system, with the $Z$ 
axis pointing in the nominal proton beam direction, referred to as the ``forward
direction'', and the $X$ axis pointing  towards the centre of HERA.
The coordinate origin is at the centre of the central tracking detector.\xspace}
\newcommand{\ZpsrapA}{%
The pseudorapidity is defined as $\eta=-\ln\left(\tan\frac{\theta}{2}\right)$,
where the polar angle, $\theta$, is measured with respect to the
$Z$ axis.\xspace}
\newcommand{\ZpsrapB}{%
The pseudorapidity is defined as $\eta=-\ln\left(\tan\frac{\theta}{2}\right)$, 
where the polar angle, $\theta$, is measured with respect to the
$Z$ axis. The azimuthal angle, $\phi$, is
measured with respect to the $X$ axis.\xspace}

\newcommand{\ZcoosysfnCEeta}{\footnote{\ZcoosysCE\ZpsrapA}}

\newcommand{\Zsttdesc}[1]{%
The STT consisted of 48 sectors of two different sizes. Each sector
contained 192 (small sector) or 264 (large sector) straws of diameter
\unit[7.5]{\mm} arranged into 3 layers. The sectors were trapezoidal in shape
and each subtended an azimuthal angle of $60\degree$ -- 6 sectors
formed a so-called superlayer. A particle passing through the complete
detector traversed 8 superlayers, which were rotated around the beam
direction at angles of $30\degree$ or $15\degree$ to each other. The STT
covered the polar-angle region $5\degree < \theta < 23\degree$.
}
\def\citeCTD{{\cite{%
nim:a279:290,*npps:b32:181,*nim:a338:254%
}}\xspace}
\def\citeMVD{{\cite{%
nim:a581:656%
}}\xspace}
\def\citeCAL{{\cite{%
nim:a309:77,*nim:a309:101,*nim:a321:356,*nim:a336:23%
}}\xspace}

\newcommand{\PYTHIA}{\textsc{Pythia}\xspace}
\newcommand{\HERWIG}{\textsc{Herwig}\xspace}

\newcommand{\ET}{\ensuremath{E_{T}}\xspace}
\newcommand{\ETjet}{\ensuremath{E_{T}^{\mathrm{jet}}}\xspace}
\newcommand{\Ejet}{\ensuremath{E^{\mathrm{jet}}}\xspace}
\newcommand{\pzjet}{\ensuremath{p_Z^{\mathrm{jet}}}\xspace}

\newcommand{\Egam}{\ensuremath{E^{\gamma}}\xspace}
\newcommand{\pzgam}{\ensuremath{p_Z^{\gamma}}\xspace}
\newcommand{\ETgam}{\ensuremath{E_{T}^{\gamma}}\xspace}
\newcommand{\etagam}{\ensuremath{\eta^{\gamma}}\xspace}
\newcommand{\etajet}{\ensuremath{\eta^{\mathrm{jet}}}\xspace}
\newcommand{\delphi}{\ensuremath{\Delta\phi}\xspace}
\newcommand{\deleta}{\ensuremath{\etagam-\etajet}\xspace}

\newcommand{\xgamm}{\ensuremath{x_{\gamma}^{\mathrm{meas}}}\xspace}
\newcommand{\xp}{\ensuremath{x_{p}^{\mathrm{obs}}}\xspace}
\newcommand{\Zacknowledge}{%
We appreciate the contributions to the construction, maintenance and
operation of the ZEUS detector made by many people who are not listed
as authors. The HERA machine group and the DESY computing staff are
especially acknowledged for their success in providing excellent
operation of the collider and the data-analysis environment. We thank
the DESY directorate for their strong support and encouragement.}

\begin{document}
\prepnum{DESY-14-086}

\title{
Further studies of the photoproduction of isolated photons with a jet at HERA
}                                                       
                    
\author{ZEUS Collaboration}
\date{\today}

\abstract{
In this extended analysis using the ZEUS detector at HERA, the
photoproduction of isolated photons together with a jet is measured
for different ranges of the fractional photon energy, \xgamm, 
contributing to the photon-jet final state.  Cross sections are
evaluated in the photon transverse-energy and pseudorapidity ranges
\mbox{$6 < E_T^{\gamma} < 15\,\gev$} and \mbox{$-0.7 < \eta^{\gamma} <
0.9$,} and for jet transverse-energy and pseudorapidity ranges $4 <
\ETjet < 35\,\gev$ and $-1.5 < \etajet < 1.8$, for an integrated
luminosity of 374 $\mathrm{pb}^{-1}$. The kinematic observables
studied comprise the transverse energy and pseudorapidity of the
photon and the jet, the azimuthal difference between them, the
fraction of proton energy taking part in the interaction, and the
difference between the pseudorapidities of the photon and the
jet. Higher-order theoretical calculations are compared to the
results. }

\makezeustitle

%
%
%
%

\def\3{\ss}
\pagenumbering{Roman}
                                                   %
\begin{center}
{                      \Large  The ZEUS Collaboration              }
\end{center}

{\small


        {\raggedright
H.~Abramowicz$^{27, u}$, 
I.~Abt$^{21}$, 
L.~Adamczyk$^{8}$, 
M.~Adamus$^{34}$, 
R.~Aggarwal$^{4, a}$, 
S.~Antonelli$^{2}$, 
O.~Arslan$^{3}$, 
V.~Aushev$^{16, 17, o}$, 
Y.~Aushev$^{17, o, p}$, 
O.~Bachynska$^{10}$
% ...
        }

\newpage


\makebox[3em]{$^{1}$}
\begin{minipage}[t]{14cm}
{\it INFN Bologna, Bologna, Italy}~$^{A}$

\end{minipage}\\
\makebox[3em]{$^{2}$}
\begin{minipage}[t]{14cm}
{\it University and INFN Bologna, Bologna, Italy}~$^{A}$

\end{minipage}\\
\makebox[3em]{$^{3}$}
\begin{minipage}[t]{14cm}
{\it Physikalisches Institut der Universit\"at Bonn,
Bonn, Germany}~$^{B}$

\end{minipage}\\
\makebox[3em]{$^{4}$}
\begin{minipage}[t]{14cm}
{\it Panjab University, Department of Physics, Chandigarh, India}

\end{minipage}\\
\makebox[3em]{$^{5}$}
\begin{minipage}[t]{14cm}
{\it Calabria University,
Physics Department and INFN, Cosenza, Italy}~$^{A}$

\end{minipage}\\
\makebox[3em]{$^{6}$}
\begin{minipage}[t]{14cm}
{\it National Centre for Particle Physics, Universiti Malaya, 50603 Kuala Lumpur, Malaysia}~$^{C}$

\end{minipage}\\
\makebox[3em]{$^{7}$}
\begin{minipage}[t]{14cm}
{\it The Henryk Niewodniczanski Institute of Nuclear Physics, Polish Academy of \\
Sciences, Krakow, Poland}~$^{D}$

\end{minipage}\\
\makebox[3em]{$^{8}$}
\begin{minipage}[t]{14cm}
{\it AGH-University of Science and Technology, Faculty of Physics and Applied Computer
Science, Krakow, Poland}~$^{D}$

\end{minipage}\\
\makebox[3em]{$^{9}$}
\begin{minipage}[t]{14cm}
{\it Department of Physics, Jagellonian University, Cracow, Poland}

\end{minipage}\\
\makebox[3em]{$^{10}$}
\begin{minipage}[t]{14cm}
{\it Deutsches Elektronen-Synchrotron DESY, Hamburg, Germany}

\end{minipage}\\
\makebox[3em]{$^{11}$}
\begin{minipage}[t]{14cm}
{\it Deutsches Elektronen-Synchrotron DESY, Zeuthen, Germany}

\end{minipage}\\
\makebox[3em]{$^{12}$}
\begin{minipage}[t]{14cm}
{\it School of Physics and Astronomy, University of Glasgow,
Glasgow, United Kingdom}~$^{E}$

\end{minipage}\\
\makebox[3em]{$^{13}$}
\begin{minipage}[t]{14cm}
{\it Hamburg University, Institute of Experimental Physics, Hamburg,
Germany}~$^{F}$

\end{minipage}\\
\makebox[3em]{$^{14}$}
\begin{minipage}[t]{14cm}
{\it Institute of Particle and Nuclear Studies, KEK,
Tsukuba, Japan}~$^{G}$

\end{minipage}\\
\makebox[3em]{$^{15}$}
\begin{minipage}[t]{14cm}
{\it Institute of Physics and Technology of Ministry of Education and
Science of Kazakhstan, Almaty, Kazakhstan}

\end{minipage}\\
\makebox[3em]{$^{16}$}
\begin{minipage}[t]{14cm}
{\it Institute for Nuclear Research, National Academy of Sciences, Kyiv, Ukraine}

\end{minipage}\\
\makebox[3em]{$^{17}$}
\begin{minipage}[t]{14cm}
{\it Department of Nuclear Physics, National Taras Shevchenko University of Kyiv, Kyiv, Ukraine}

\end{minipage}\\
\makebox[3em]{$^{18}$}
\begin{minipage}[t]{14cm}
{\it Department of Physics, McGill University,
Montr\'eal, Qu\'ebec, Canada H3A 2T8}~$^{H}$

\end{minipage}\\
\makebox[3em]{$^{19}$}
\begin{minipage}[t]{14cm}
{\it Meiji Gakuin University, Faculty of General Education,
Yokohama, Japan}~$^{G}$

\end{minipage}\\
\makebox[3em]{$^{20}$}
\begin{minipage}[t]{14cm}
{\it Lomonosov Moscow State University, Skobeltsyn Institute of Nuclear Physics,
Moscow, Russia}~$^{I}$

\end{minipage}\\
\makebox[3em]{$^{21}$}
\begin{minipage}[t]{14cm}
{\it Max-Planck-Institut f\"ur Physik, M\"unchen, Germany}

\end{minipage}\\
\makebox[3em]{$^{22}$}
\begin{minipage}[t]{14cm}
{\it Department of Physics, University of Oxford,
Oxford, United Kingdom}~$^{E}$

\end{minipage}\\
\makebox[3em]{$^{23}$}
\begin{minipage}[t]{14cm}
{\it INFN Padova, Padova, Italy}~$^{A}$

\end{minipage}\\
\makebox[3em]{$^{24}$}
\begin{minipage}[t]{14cm}
{\it Dipartimento di Fisica e Astronomia dell' Universit\`a and INFN,
Padova, Italy}~$^{A}$

\end{minipage}\\
\makebox[3em]{$^{25}$}
\begin{minipage}[t]{14cm}
{\it Polytechnic University, Tokyo, Japan}~$^{G}$

\end{minipage}\\
\makebox[3em]{$^{26}$}
\begin{minipage}[t]{14cm}
{\it Dipartimento di Fisica, Universit\`a `La Sapienza' and INFN,
Rome, Italy}~$^{A}$

\end{minipage}\\
\makebox[3em]{$^{27}$}
\begin{minipage}[t]{14cm}
{\it Raymond and Beverly Sackler Faculty of Exact Sciences, School of Physics, \\
Tel Aviv University, Tel Aviv, Israel}~$^{J}$

\end{minipage}\\
\makebox[3em]{$^{28}$}
\begin{minipage}[t]{14cm}
{\it Department of Physics, Tokyo Institute of Technology,
Tokyo, Japan}~$^{G}$

\end{minipage}\\
\makebox[3em]{$^{29}$}
\begin{minipage}[t]{14cm}
{\it Universit\`a di Torino and INFN, Torino, Italy}~$^{A}$

\end{minipage}\\
\makebox[3em]{$^{30}$}
\begin{minipage}[t]{14cm}
{\it Universit\`a del Piemonte Orientale, Novara, and INFN, Torino,
Italy}~$^{A}$

\end{minipage}\\
\makebox[3em]{$^{31}$}
\begin{minipage}[t]{14cm}
{\it Department of Physics, University of Toronto, Toronto, Ontario,
Canada M5S 1A7}~$^{H}$

\end{minipage}\\
\makebox[3em]{$^{32}$}
\begin{minipage}[t]{14cm}
{\it Physics and Astronomy Department, University College London,
London, United Kingdom}~$^{E}$

\end{minipage}\\
\makebox[3em]{$^{33}$}
\begin{minipage}[t]{14cm}
{\it Faculty of Physics, University of Warsaw, Warsaw, Poland}

\end{minipage}\\
\makebox[3em]{$^{34}$}
\begin{minipage}[t]{14cm}
{\it National Centre for Nuclear Research, Warsaw, Poland}

\end{minipage}\\
\makebox[3em]{$^{35}$}
\begin{minipage}[t]{14cm}
{\it Department of Particle Physics and Astrophysics, Weizmann
Institute, Rehovot, Israel}

\end{minipage}\\
\makebox[3em]{$^{36}$}
\begin{minipage}[t]{14cm}
{\it Department of Physics, York University, Ontario, Canada M3J 1P3}~$^{H}$

\end{minipage}\\


\makebox[3ex]{$^{ A}$}
\begin{minipage}[t]{14cm}
 supported by the Italian National Institute for Nuclear Physics (INFN) \
\end{minipage}\\
\makebox[3ex]{$^{ B}$}
\begin{minipage}[t]{14cm}
 supported by the German Federal Ministry for Education and Research (BMBF), under
 contract No. 05 H09PDF\
\end{minipage}\\
\makebox[3ex]{$^{ C}$}
\begin{minipage}[t]{14cm}
 supported by HIR grant UM.C/625/1/HIR/149 and UMRG grants RU006-2013, RP012A-13AFR and RP012B-13AFR from
 Universiti Malaya, and ERGS grant ER004-2012A from the Ministry of Education, Malaysia\
\end{minipage}\\
\makebox[3ex]{$^{ D}$}
\begin{minipage}[t]{14cm}
 supported by the National Science Centre under contract No. DEC-2012/06/M/ST2/00428\
\end{minipage}\\
\makebox[3ex]{$^{ E}$}
\begin{minipage}[t]{14cm}
 supported by the Science and Technology Facilities Council, UK\
\end{minipage}\\
\makebox[3ex]{$^{ F}$}
\begin{minipage}[t]{14cm}
 supported by the German Federal Ministry for Education and Research (BMBF), under
 contract No. 05h09GUF, and the SFB 676 of the Deutsche Forschungsgemeinschaft (DFG) \
\end{minipage}\\
\makebox[3ex]{$^{ G}$}
\begin{minipage}[t]{14cm}
 supported by the Japanese Ministry of Education, Culture, Sports, Science and Technology
 (MEXT) and its grants for Scientific Research\
\end{minipage}\\
\makebox[3ex]{$^{ H}$}
\begin{minipage}[t]{14cm}
 supported by the Natural Sciences and Engineering Research Council of Canada (NSERC) \
\end{minipage}\\
\makebox[3ex]{$^{ I}$}
\begin{minipage}[t]{14cm}
 supported by RF Presidential grant N 3042.2014.2 for the Leading Scientific Schools and by
 the Russian Ministry of Education and Science through its grant for Scientific Research on
 High Energy Physics\
\end{minipage}\\
\makebox[3ex]{$^{ J}$}
\begin{minipage}[t]{14cm}
 supported by the Israel Science Foundation\
\end{minipage}\\
\vspace{30em} \pagebreak[4]


\makebox[3ex]{$^{ a}$}
\begin{minipage}[t]{14cm}
also funded by Max Planck Institute for Physics, Munich, Germany\
\end{minipage}\\
\makebox[3ex]{$^{ b}$}
\begin{minipage}[t]{14cm}
also funded by Max Planck Institute for Physics, Munich, Germany, now at Sri Guru Granth Sahib World University, Fatehgarh Sahib\
\end{minipage}\\
\makebox[3ex]{$^{ c}$}
\begin{minipage}[t]{14cm}
also at Agensi Nuklear Malaysia, 43000 Kajang, Bangi, Malaysia\
\end{minipage}\\
\makebox[3ex]{$^{ d}$}
\begin{minipage}[t]{14cm}
also at Cracow University of Technology, Faculty of Physics, Mathematics and Applied Computer Science, Poland\
\end{minipage}\\
\makebox[3ex]{$^{ e}$}
\begin{minipage}[t]{14cm}
partially supported by the Polish National Science Centre projects DEC-2011/01/B/ST2/03643 and DEC-2011/03/B/ST2/00220\
\end{minipage}\\
\makebox[3ex]{$^{ f}$}
\begin{minipage}[t]{14cm}
now at Rockefeller University, New York, NY 10065, USA\
\end{minipage}\\
\makebox[3ex]{$^{ g}$}
\begin{minipage}[t]{14cm}
now at University of Liverpool, United Kingdom\
\end{minipage}\\
\makebox[3ex]{$^{ h}$}
\begin{minipage}[t]{14cm}
also affiliated with University College London, UK\
\end{minipage}\\
\makebox[3ex]{$^{ i}$}
\begin{minipage}[t]{14cm}
supported by the Alexander von Humboldt Foundation\
\end{minipage}\\
\makebox[3ex]{$^{ j}$}
\begin{minipage}[t]{14cm}
now at CERN, Geneva, Switzerland\
\end{minipage}\\
\makebox[3ex]{$^{ k}$}
\begin{minipage}[t]{14cm}
also at Institute of Theoretical and Experimental Physics, Moscow, Russia\
\end{minipage}\\
\makebox[3ex]{$^{ l}$}
\begin{minipage}[t]{14cm}
Alexander von Humboldt Professor; also at DESY and University of Oxford\
\end{minipage}\\
\makebox[3ex]{$^{ m}$}
\begin{minipage}[t]{14cm}
also at University of Tokyo, Japan\
\end{minipage}\\
\makebox[3ex]{$^{ n}$}
\begin{minipage}[t]{14cm}
now at Kobe University, Japan\
\end{minipage}\\
\makebox[3ex]{$^{ o}$}
\begin{minipage}[t]{14cm}
supported by DESY, Germany\
\end{minipage}\\
\makebox[3ex]{$^{ p}$}
\begin{minipage}[t]{14cm}
member of National Technical University of Ukraine, Kyiv Polytechnic Institute, Kyiv, Ukraine\
\end{minipage}\\
\makebox[3ex]{$^{ q}$}
\begin{minipage}[t]{14cm}
now at DESY ATLAS group\
\end{minipage}\\
\makebox[3ex]{$^{ r}$}
\begin{minipage}[t]{14cm}
member of National University of Kyiv - Mohyla Academy, Kyiv, Ukraine\
\end{minipage}\\
\makebox[3ex]{$^{ s}$}
\begin{minipage}[t]{14cm}
now at BNL, USA\
\end{minipage}\\
\makebox[3ex]{$^{ t}$}
\begin{minipage}[t]{14cm}
now at LNF, Frascati, Italy\
\end{minipage}\\
\makebox[3ex]{$^{ u}$}
\begin{minipage}[t]{14cm}
also at Max Planck Institute for Physics, Munich, Germany, External Scientific Member\
\end{minipage}\\
\makebox[3ex]{$^{ v}$}
\begin{minipage}[t]{14cm}
now at Tokyo Metropolitan University, Japan\
\end{minipage}\\
\makebox[3ex]{$^{ w}$}
\begin{minipage}[t]{14cm}
also supported by DESY\
\end{minipage}\\
\makebox[3ex]{$^{ x}$}
\begin{minipage}[t]{14cm}
also at \L\'{o}d\'{z} University, Poland\
\end{minipage}\\
\makebox[3ex]{$^{ y}$}
\begin{minipage}[t]{14cm}
member of \L\'{o}d\'{z} University, Poland\
\end{minipage}\\
\makebox[3ex]{$^{ z}$}
\begin{minipage}[t]{14cm}
now at Polish Air Force Academy in Deblin\
\end{minipage}\\
\makebox[3ex]{$^{aa}$}
\begin{minipage}[t]{14cm}
deceased\
\end{minipage}\\
\pagebreak[4]
}

\pagenumbering{arabic} 
\pagestyle{plain}

\section{Introduction}
\label{sec-int}

In a recently published paper~\cite{pl:b730:293}, the ZEUS
collaboration presented cross sections for events containing an
isolated high-energy photon, with and without a jet, produced in
photoproduction at the HERA collider using the full HERA II data set.
Such events can provide a direct probe of the underlying partonic
process in high-energy collisions involving photons, since the
emission of a high-energy photon is largely unaffected by parton
hadronisation.  In photoproduction processes  in $ep$ collisions at HERA, the
exchanged virtual photon is quasi-real, with small virtuality,
$Q^2$, conventionally required to be less than 1~GeV$^2$. These measurements follow earlier analyses of isolated photons
in photoproduction by the ZEUS and H1 collaborations
\cite{pl:b413:201,pl:b472:175,pl:b511:19,epj:c49:511,epj:c38:437,epj:c66:17}, 
as well as in deep inelastic scattering
(DIS)~\cite{pl:b595:86,epj:c54:371,pl:b687:16,pl:b715:88}.  In the 
analysis presented here, the most recent ZEUS photoproduction measurements
are extended, using the same data as used previously.

\begin{figure}
\vfill
\begin{center}
\epsfig{file=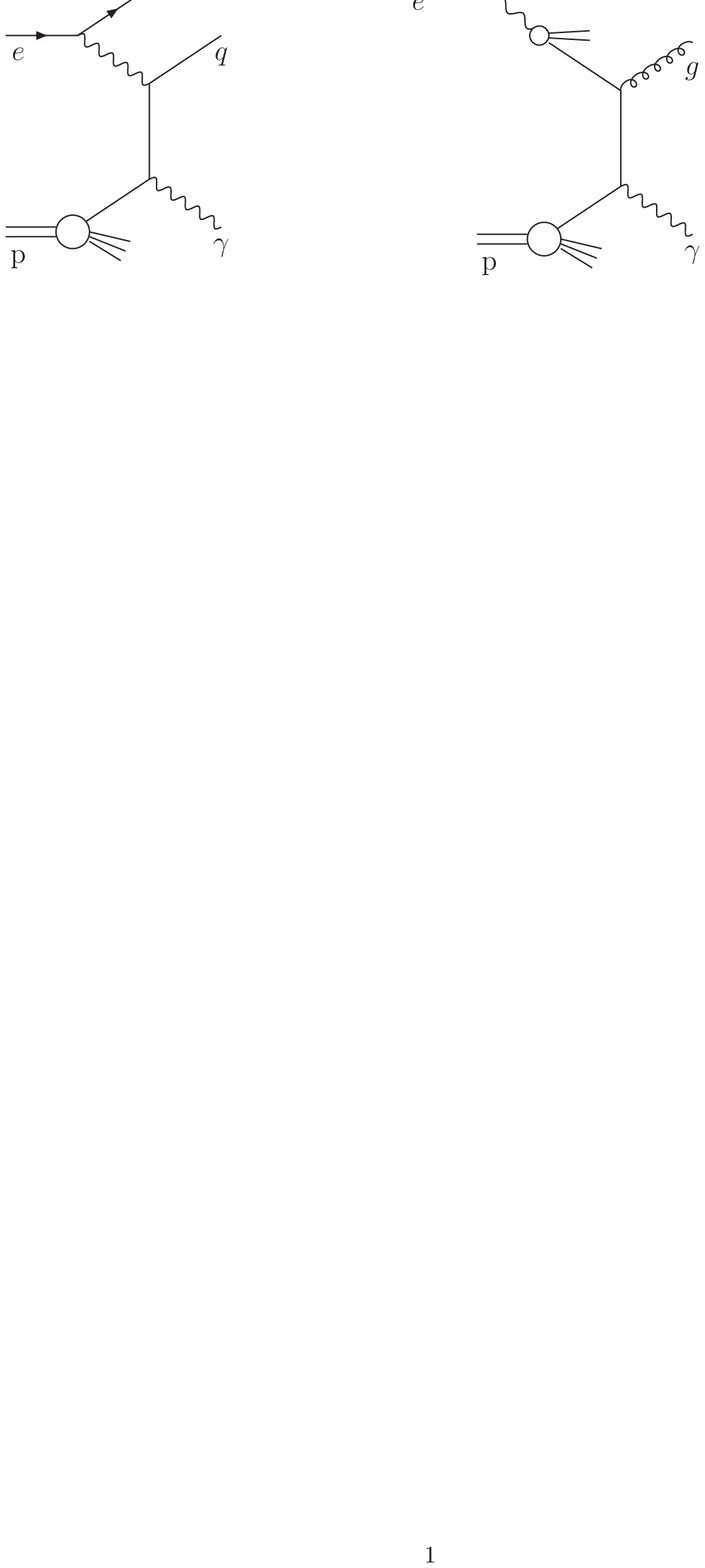,width=11cm,%
bbllx=120,bblly=670,bburx=420,bbury=820}
\hspace*{10mm}
\\[-4mm]
(a)\hspace{7.5cm}(b)\\[5mm]
\epsfig{file=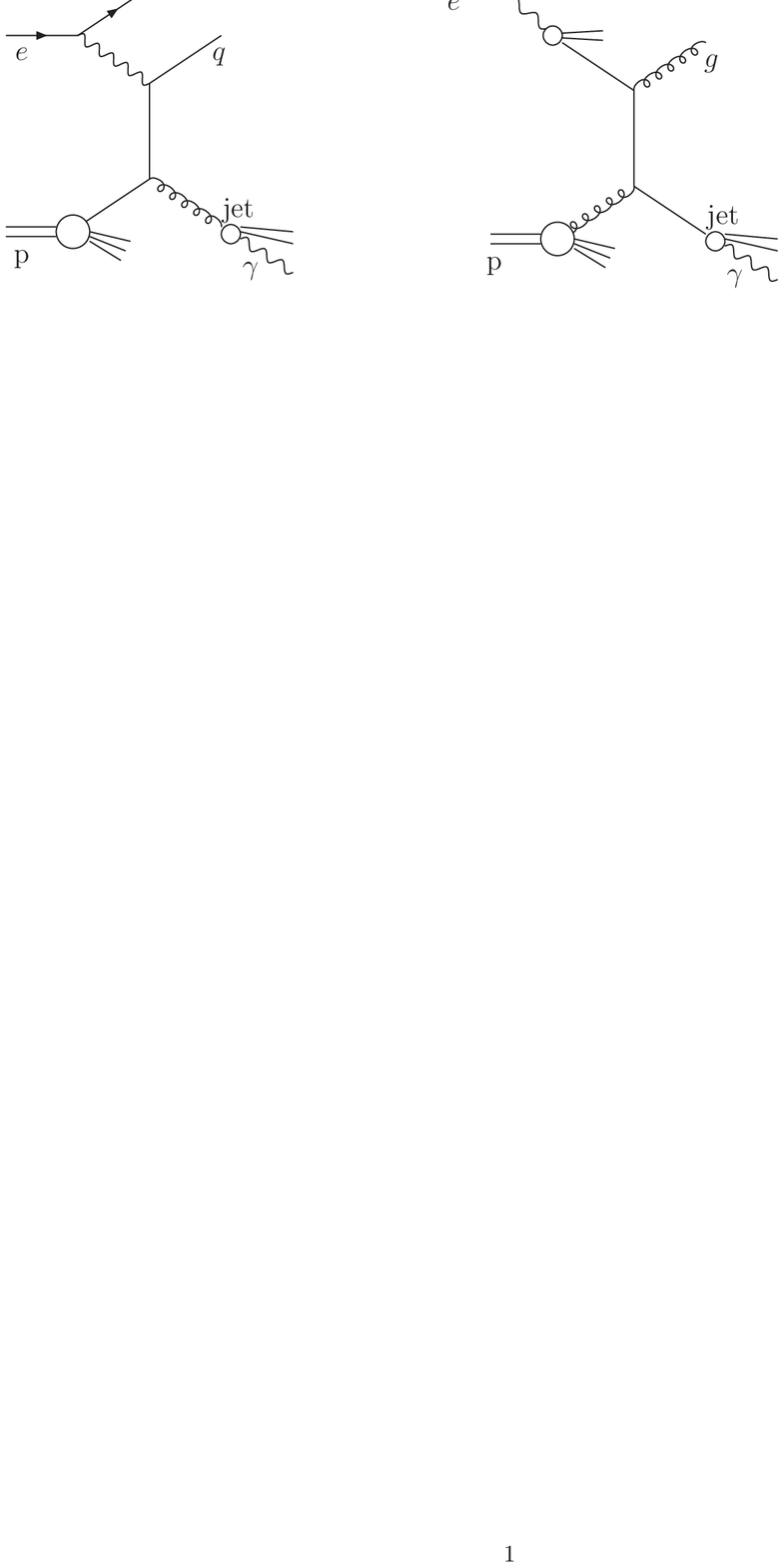,width=12cm,%
bbllx=90,bblly=670,bburx=420,bbury=820}
\\[-4mm]
(c)\hspace{7.5cm}(d)
\end{center}
\vspace*{0mm}
\caption{Examples of (a) direct-prompt and (b) resolved-prompt 
processes at leading order in QCD, and the related (c) direct and 
(d) resolved fragmentation processes.}
\label{fig1}
\vfill
\end{figure}

In ``direct'' production processes, the entire incoming photon is
absorbed by an outgoing quark from the incoming proton, while in
``resolved'' processes, the photon's hadronic structure provides a
quark or gluon that interacts with a parton from the proton.
Figure~\ref{fig1} gives examples of the lowest-order (LO) direct and
resolved diagrams for high-energy photoproduction of photons in
quantum chromodynamics (QCD)\footnote{Photons that are radiated in the
hard scattering process, rather than resulting from meson decay, are
commonly called ``prompt''. An alternative nomenclature
is to refer to such photons as ``direct''; thus Figs.~\ref{fig1}(a)
and \ref{fig1}(b) would be called ``direct-direct'' and
``resolved-direct'' diagrams, respectively.}.  Higher-order processes
also include ``fragmentation processes'' in which a photon is radiated
within a jet, also illustrated in Fig.~\ref{fig1}. Such processes are
suppressed by requiring that the outgoing photon must be isolated.

Resolved and direct processes may be
partially distinguished in events containing a high-\ET\ photon and a jet
by means of the quantity
\begin{equation}
\xgamm =  \frac{\Egam+\Ejet-\pzgam-\pzjet}
{E^{\text{all}}- p_Z^{\text{all}}}, 
\end{equation}
which measures the fraction of the incoming photon energy that is
given to the photon and the jet.  The quantities \Egam\ and \Ejet
denote the energies of the photon and the jet, respectively, $p_Z$
denotes the corresponding longitudinal momenta{\ZcoosysfnCEeta}, and
the suffix ``all'' refers to all the measured final-state particles of
an event. At LO, \xgamm= 1 for direct events, while any value in the
range (0,1) may be taken for resolved events. At higher order, the
first statement no longer precisely holds, but the presence of
direct processes generates a prominent peak in the cross section at
high \xgamm. Here, measurements in a direct-dominated region are
presented by selecting events with \xgamm$>0.8$, and in a
resolved-dominated region by selecting events with \xgamm$<0.8$.  This
enables the behaviour of the photoproduction process to be explored in
more detail.

Several kinematic quantities are also measured beyond those presented
previously.   The quantity
$$\xp = (\ETgam\exp{\etagam} +\ETjet\exp{\etajet})/2E_p$$ estimates
the fraction of proton energy taken by the parton that interacts with
the photon; its distribution is sensitive to the proton's partonic
structure.  Here,
\ET\ denotes transverse energy, $\eta$ denotes pseudorapidity, and
$E_p$ is the energy of the proton beam.  The difference in
pseudorapidities, \deleta, is sensitive to the dynamical details of
the hard scattering process, in particular to the spin of the
exchanged quantum~\cite{pl:b384:401}. The quantity \delphi, defined as
the absolute difference between the azimuths of the photon and the
high-\ET\ jet, is sensitive to the presence of higher-order gluon
radiation in the event, especially relative to the outgoing quark,
which can generate non-collinearity between the photon and the leading
jet. All three of these quantities are insensitive to Lorentz boosts
along the $Z$ axis.

Predictions from QCD-based models are compared to the measurements.  The
cross sections for isolated-photon production in photoproduction have
been calculated to next-to-leading order (NLO) by Fontannaz,
Guillet and Heinrich (FGH)~\cite{epj:c21:303,epj:c34:191,priv:fgh:2013}.  
Calculations based on the $k_T$-factorisation approach have been made by
Lipatov, Malyshev and Zotov 
(LMZ)~\cite{pr:d72:054002,pr:d81:094027,pr:d88:074001,lmzpriv}.

\section{Experimental set-up}
\label{sec-exp}
The measurements are based on a data sample corresponding to an
integrated luminosity of $374\pm 7 \,\mathrm{pb}^{-1}$, taken during
the years 2004 to 2007 with the ZEUS detector at HERA. During this
period, HERA ran with an electron or positron beam energy of 27.5\gev\
and a proton beam energy of $E_p =920$\gev. The sample is a sum of
$e^+p$ and $e^-p$ data\footnote{Hereafter ``electron'' refers to both
electrons and positrons unless otherwise stated.}.

A detailed description of the ZEUS detector can be found
elsewhere~\cite{zeus:1993:bluebook}. Charged particles were measured in
the central tracking detector (CTD)~\citeCTD and a silicon micro
vertex detector (MVD)~\cite{nim:a581:656} which operated in a magnetic
field of $1.43$~T provided by a thin superconducting solenoid.  The
high-resolution uranium--scintillator calorimeter (CAL)~\citeCAL
consisted of three parts: the forward (FCAL), the barrel (BCAL) and
the rear (RCAL) calorimeters. The BCAL covered the pseudorapidity
range $-0.74$ to 1.01 as seen from the nominal interaction point, and the
FCAL and RCAL extended the coverage to the range $-3.5$ to 4.0.  Each
part of the CAL was subdivided into elements referred to as cells. The
barrel electromagnetic calorimeter (BEMC) cells had a pointing
geometry aimed at the nominal interaction point, with a cross section
approximately $5\times20\,\mathrm{cm^2}$, with the finer granularity
in the $Z$ direction and the coarser in the $(X,Y)$
plane.  This fine granularity allows the use of shower-shape
distributions to distinguish isolated photons from the products of
neutral meson decays such as $\pi^0 \rightarrow \gamma\gamma$.
The CAL energy resolution, as measured under test-beam conditions, was
$\sigma(E)/E = 0.18/\sqrt{E}$ for electrons and $0.35/\sqrt{E}$
for hadrons, where $E$ is in \gev.  

The luminosity was measured \cite{lumi2} using the Bethe--Heitler reaction $ep
\rightarrow e\gamma p$ by a luminosity detector which consisted of two
independent systems: a lead--scintillator calorimeter
\cite{desy-92-066,*zfp:c63:391,*acpp:b32:2025} and a magnetic
spectrometer~\cite{nim:a565:572}.


\section{Theoretical models}
\label{sec:theory}

Two theoretical models are considered. In the approach of
FGH~\cite{epj:c21:303,epj:c34:191}, the LO and NLO diagrams and the
box-diagram term are calculated explicitly. Fragmentation processes
are calculated in terms of a fragmentation function in which a quark
or gluon gives rise to a photon; an experimentally determined
non-perturbative parameterisation is used as input to the theoretical
calculation~\cite{epj:c19:89}.  Fragmentation and box terms each contribute
about 10\% to the total cross section.  The
CTEQ6~\cite{jhep:0602:032} and AFG04~\cite{epj:c44:395} parton
densities are used for the proton and photon, respectively. Theoretical
uncertainties arise due to the choice of renormalisation,
factorisation and fragmentation scales. They were estimated, using a
more conservative approach~\cite{priv:fgh:2013} than in the original
published paper~\cite{epj:c21:303}, by varying the renormalisation
scale by factors of 0.5 and 2.0, since this gave the largest effect on
the cross sections.

The $k_T$-factorisation method used by
LMZ~\cite{pr:d72:054002,pr:d81:094027,pr:d88:074001} makes use of
unintegrated parton densities in the proton, using the KMR
formalism~\cite{kmr} based on the MSTW08 proton parton
densities~\cite{epj:c63:189}. In addition to the hard QCD subprocess,
the model incorporates a parton evolution cascade, one jet from which
can be taken as the leading jet in the analysis.
Fragmentation terms and the quark content of the resolved photon are
not included, but the box diagram is included together with $2\to3$
subprocesses to represent the LO direct and resolved photon
contributions. The calculation used in the previous ZEUS analysis~\cite{pl:b730:293} has been
augmented by a term that takes account of the gluon content of the
resolved photon, and further technical changes have been
implemented~\cite{lmzpriv}.  Uncertainties associated with the hard
scale were provided by the authors.  There is a further overall
statistical uncertainty on the set of results for each variable, of
the order of 10\% for the results presented here.

All results are presented at the hadron level;  
to make use of the theoretical predictions, cuts equivalent to the experimental
kinematic selections including the photon isolation (see Section
\ref{sec-selec}) were applied at the parton level.  Hadronisation
corrections were then evaluated (Section \ref{sec-mc}) and applied to
the theoretical calculations to enable a comparison to the
experimental data.


\section{Monte Carlo event simulation}
\label{sec-mc}
Monte Carlo (MC) event samples were employed to evaluate the detector
acceptance and event-reconstruction efficiency, and to provide signal
and background distributions. The program \PYTHIA
6.416~\cite{jhep:0605:026} was used to generate the direct and
resolved prompt-photon processes at LO, and also $2\to2$ parton-parton
scattering processes not involving photons (``dijet events''), making
use of the CTEQ4~\cite{pr:d55:1280} and GRV~\cite{grv} proton and
photon parton densities.  The program was run using the default
parameters with minor modifications\footnote{In particular, the
\PYTHIA\ parameter {\sc parp(67)} was set to 4.0 and multiple parton
interactions were turned off.  In \HERWIG the parameters {\sc ispac,
qspac,} and {\sc ptrms} were set to 2, 4.0, and 0.44.}.  The isolated
photons measured in the experiment are accompanied by backgrounds from
neutral mesons in hadronic jets, in particular $\pi^0$ and $\eta$,
where the meson decay products create an energy cluster in the BCAL
that passes the selection criteria for a photon.  The dijet event
samples included background events of this kind which were extracted
for use in the analysis.  The \PYTHIA dijet events in which a
high-energy photon was radiated from a quark or lepton (``radiative
events'') were not used in the background samples but were
defined, in accordance with theory, as a component of the signal.

Event samples were also generated using the \HERWIG 6.510
program~\cite{jhep:0101:010}, again with minor modifications to the
default parameters.  The \PYTHIA and \HERWIG programs differ
significantly in their treatment of parton showers, and in the use of
a string-based hadronisation scheme in \PYTHIA but a cluster-based 
scheme in \HERWIG.

The generated MC events were passed through the ZEUS detector and
trigger simulation programs based on {\sc Geant}
3.21~\cite{tech:cern-dd-ee-84-1}. They were then reconstructed and analysed
using the same programs as used for the data.  The hadronisation corrections to
the theory calculations were evaluated using \PYTHIA and
\HERWIG, the two programs being in agreement to a few percent;
\PYTHIA was used to provide the values for the present analysis.
No uncertainties were applied to these corrections.  They were
calculated by running the same jet algorithm and event selections,
including the isolation criterion, on the generated partons and on the
hadronised final state in the direct and resolved prompt-photon MC
events.


\section{Event selection and reconstruction}
\label{sec-selec}

The basic event selection and reconstruction was performed as
previously.  A three-level trigger system was used
to select events online
\cite{zeus:1993:bluebook,uproc:chep:1992:222,nim:a580:1257}:
\begin{itemize}
\item the first-level trigger required a loosely measured track in the CTD
and a minimum energy deposited in the CAL;
\item at the second level, the event conditions were tightened;
\item at the third level, the event was reconstructed and a high-energy 
photon candidate was required.
Most deep inelastic scattering events were rejected.
\end{itemize}

In the offline event analysis, some general conditions were applied as
follows:
\begin{itemize}
\item to reduce background from non-$ep$
collisions, events were required to have a reconstructed vertex
position, $Z_{\mathrm{vtx}}$, within the range $|Z_{\mathrm{vtx}}|<
40\,\mathrm{cm}$;  
\item to remove any DIS contamination, no scattered beam electron was permitted in the
ZEUS detector;
\item a range of incoming virtual photon energies was selected 
by the requirement
$0.2 < y_{\JB} < 0.7$, where $y_{\JB} = \sum \limits_i E_i(1-\cos
\theta_i)/2E_e$ and $E_e$ is the energy of the electron beam.  
Here, $E_i$ is the energy of the $i$-th CAL cell, $\theta_i$ is
its polar angle and the sum runs over all cells~\cite{pl:b303:183}.
The lower cut strengthened the trigger requirements and the upper cut
further suppressed remaining deep inelastic scattering events. 
\end{itemize}

The subsequent event analysis made use of energy-flow objects
(EFOs)~\cite{epj:c1:81,*epj:c6:43}, which  were constructed
from clusters of calorimeter cells, associated with
tracks when appropriate. Tracks not associated with calorimeter
clusters were also used.  Photon candidates were 
EFOs with no associated track and with at least $90\%$ of the
reconstructed energy measured in the BEMC. Candidate EFOs with wider
electromagnetic showers than are typical for a single photon were
accepted, in order to evaluate the backgrounds. 

Jet reconstruction was performed, making use of all the EFOs in the
event including photon candidates, by means of the $k_T$ clustering
algorithm~\cite{np:b406:187} in the $E$-scheme in the longitudinally
invariant inclusive mode~\cite{pr:d48:3160} with the radius parameter
set to 1.0.  By construction, one of the jets found by this procedure
corresponds to or includes the photon candidate. An additional
accompanying jet was required; if more than one was found in the
designated angular range, that with the highest transverse energy,
$E_T^{\mathrm{jet}},$ was used.  In the kinematic region used, the
resolution of the jet transverse energy was about 15--20\%, estimated
using MC simulations.

To reduce the fragmentation contribution and the background from the
decay of neutral mesons within jets, the photon candidate was required
to be isolated from other hadronic
activity.  This was imposed by requiring that the photon-candidate EFO
had at least 90\% of the total energy of the reconstructed jet of
which it formed a part. High-\ET\ photons radiated from scattered
leptons were further suppressed by rejecting photons with a near-by
track. This was achieved by demanding $\Delta R > 0.2,$ where $\Delta
R = \sqrt{(\Delta \phi)^2 + (\Delta\eta)^2}$ is the distance to the
nearest reconstructed track with momentum greater than
$250\,\mathrm{MeV}$ in the $\eta \-- \phi$ plane, where $\phi$ is the
azimuthal angle. This latter condition was applied only at the
detector level, and not in the hadron- or parton-level calculations.

Events were finally selected with the following kinematic conditions:
\begin{itemize}
\item each event was required to contain an isolated photon candidate with a 
reconstructed transverse energy, $E_T^{\gamma}$, in the range
\mbox{$6 <E_T^{\gamma}<15\,\mathrm{\gev\ }$} and with pseudorapidity,
$\eta^{\gamma}$, in the range $-0.7 < \eta^{\gamma} < 0.9$;  
\item  a hadronic jet was required with 
$E_T^{\mathrm{jet}}$ between 4 and 35\gev\ and lying within the
pseudorapidity, $\eta^{\mathrm{jet}}$, range $-1.5
<\eta^{\mathrm{jet}} < 1.8$;
\item selections were made for all \xgamm,
giving a total of 12450 events, and for $\xgamm>0.8$ and $\xgamm<0.8$.
The latter two conditions selected events in direct-enhanced and
resolved-enhanced regions, respectively.  An additional selection was made
with events having $\xgamm<0.7$.
\end{itemize} 

\section{Extraction of the photon signal}

The selected samples contain a large admixture of background
events in which one or more neutral mesons, such as $\pi^0$ and
$\eta$, have decayed to photons, thereby producing a photon candidate in
the BEMC.  The photon signal was extracted statistically following the 
approach used in previous ZEUS analyses
\cite{pl:b595:86,epj:c54:371,pl:b687:16,pl:b715:88,pl:b730:293}.  
The method made use of the
energy-weighted width, measured in the $Z$ direction, of the BEMC
energy-cluster comprising the photon candidate. This width was calculated
as $$\langle\delta Z\rangle=
\sum \limits_i E_i|Z_i-Z_{\mathrm{cluster}}|
{\large/}( w_{\mathrm{cell}}\sum \limits_i E_i),$$ where $Z_{i}$ is
the $Z$ position of the centre of the $i$-th cell,
$Z_{\mathrm{cluster}}$ is the energy-weighted centroid of the EFO
cluster, $w_{\mathrm{cell}}$ is the width of the cell in the $Z$
direction, and $E_i$ is the energy recorded in the cell. The sum runs
over all BEMC cells in the EFO.  

\begin{figure}

\vfill
\begin{center}
\epsfig{file=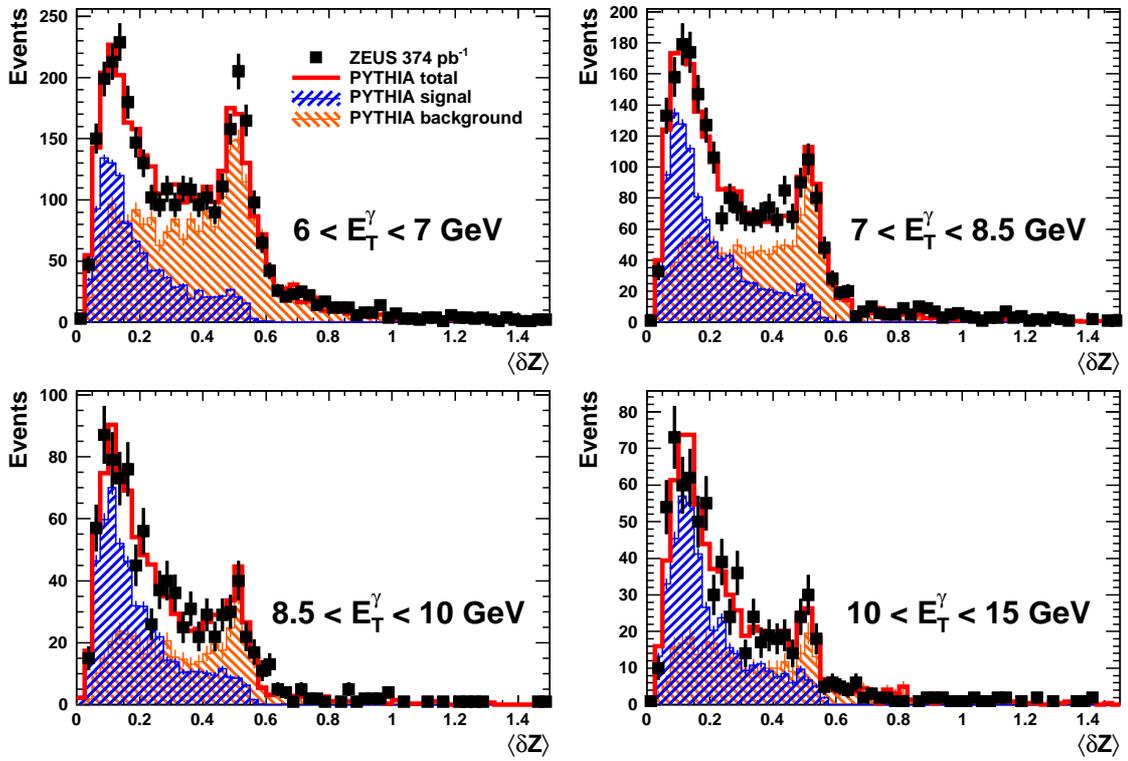,width=15cm}
\end{center}
\caption{\small 
Examples of fits to $\langle \delta Z\rangle$ for different ranges
of the photon transverse energy, showing the signal and background
contributions and the fitted total. 
}
\label{fig:zfits}
\end{figure}

The number of isolated-photon events in the data was determined by a
$\chi^2$ fit to the $\langle \delta Z \rangle$ distribution in the
range $0.05<\langle \delta Z \rangle < 0.8$, varying the relative
fractions of the signal and background components as represented by
histogram templates obtained from the MC.  The fit was performed for
each measured cross-section bin, with $\chi^2$ values of typically 1.1
per degree of freedom (e.g.~31/28), verifying that the signal and
background were well understood.  The extracted signals corresponded
overall to $6262\pm 132$ events with a photon and an accompanying jet.
A set of typical fits for different ranges of the photon transverse
energy is shown in Fig.~\ref{fig:zfits} and illustrates how the
signal-to-background ratio improves with increasing $E_T^\gamma$.

A bin-by-bin correction method was used to determine the
production cross section in a given variable, by means of the relationship

\begin{equation}
\frac{d\sigma}{dY} = \frac{\mathcal{A}\, N(\gamma)}{  
\mathcal{L} \, \Delta Y},
\end{equation}

where $N(\gamma)$ is the number of photons in a bin as extracted from
the fit, $\Delta Y$ is
the bin width, $\mathcal{L}$ is the total integrated luminosity, and
$\mathcal{A}$ is the acceptance correction.  The acceptance correction
was calculated, using MC samples, as the ratio of the number of events
that were generated in the given bin to the number of events obtained
in the bin after event reconstruction.  Its value was typically 1.2.

Allowance must be made for the different
acceptances for the direct and the resolved processes, as modelled by
\PYTHIA.  Over the entire \xgamm\ range, a reasonable phenomenological
description of the data can be obtained using a MC sample consisting
of a 50:40 mixture of \PYTHIA-simulated direct and resolved events,
normalised to the data, with a 10\% admixture of radiative events
divided equally between direct and resolved. The acceptance factors
were calculated using this model, applying selections for
the chosen \xgamm\ regions. Small corrections were applied for the
trigger efficiency modelling and a residual contamination by DIS
events~\cite{pl:b730:293}.

The photon energy scale was calibrated by means of an analysis of
deeply virtual Compton scattering events recorded by ZEUS, in which the
detected final-state particles comprised a scattered electron, whose
energy measurement is well understood, and a balancing outgoing
photon.

\section{Systematic uncertainties}
\label{sec:syst}

The most significant sources of systematic uncertainty were evaluated
as follows:

\begin{itemize}

\item to allow for uncertainties in the simulation of the 
hadronic final state, the cross sections were recalculated using
\HERWIG to model the signal and background events.  The ensuing
  changes in the results correspond to an uncertainty of typically up
  to 8\%, but rising to 18\% in the highest bin of \xgamm;

\item the energy of the photon candidate was varied by $\pm 2\%$ in the 
MC at the detector level.  Independently, the energy of the
accompanying jet was varied by an amount decreasing
from $\pm4.5\%$ to $\pm2.5\%$ as \ETjet\ increases from 4\gev\ to
above 10\gev. Each of these gave variations in the measured cross
sections of typically 5\%.
\end{itemize}

Further systematic uncertainties were evaluated as follows:

\begin{itemize}

\item the uncertainty in the acceptance due to the estimation of the 
relative fractions of direct and resolved events and radiative events
in the MC sample was estimated by varying these fractions by $\pm
15\%$ and $\pm 5\%$ respectively in absolute terms; the changes in the
cross sections were typically $\pm2\%$ in each case;

\item  the dependence of the result on the modelling of the hadronic 
background by the MC was investigated by varying the upper limit for
the $\langle
\delta Z\rangle$ fit in the range $[0.6, 1.0]$; this gave a $\pm
2\%$ variation;

\item the \ETgam, \etajet\ and \delphi\ distributions in the MC 
were reweighted simultaneously to provide a closer agreement with the
data, and the cross sections were re-evaluated.  This generated 
changes of typically $\pm$2\%.

\end{itemize}

Other sources of systematic uncertainty were found to be negligible. These included the modelling of the track-isolation
cut and the track-momentum cut, and also the cuts on photon isolation, the
electromagnetic fraction of the photon shower, $y_{\JB}$ and
$Z_{\text{vtx}}$.  Except for the uncertainty on the modelling of the
hadronic final state, the major uncertainties were treated as
symmetric, and all the uncertainties were combined in quadrature.  The
uncertainties of 2.0\% on the trigger efficiency and 1.9\% on
the luminosity measurement were not included in the tables and figures.

\section{Results}

\label{sec:results}

Differential cross sections were calculated for the production of an
isolated photon with at least one accompanying jet, in the kinematic
region defined by $Q^2< 1 \, \mathrm{\gev}^2$, $0.2 < y < 0.7,$ $-0.7<
\etagam < 0.9$, $6 < \ETgam< 15\,\mathrm{\gev} $, $4 < \ETjet <
35$\gev\ and $-1.5 <\etajet< 1.8$.  All quantities were evaluated at
the hadron level in the laboratory frame, and $y$ is defined as the
fraction of the incoming lepton energy that is given to the virtual
photon. The jets were formed according to the $k_T$ clustering
algorithm with the radius parameter set to 1.0.  Photon isolation was
imposed such that at least $90 \%$ of the energy of the jet-like
object containing the photon originated from the photon.  If more than
one accompanying jet was found within the designated \etajet range in
an event, that with highest \ETjet was taken.  Cross sections in
$\ETjet$ above 15\gev\ are omitted from the tables and
Fig.~\ref{fig:jet} owing to limited statistics, but this kinematic
region is included in the other cross-section measurements.

\begin{figure}
\vfill
\epsfig{file=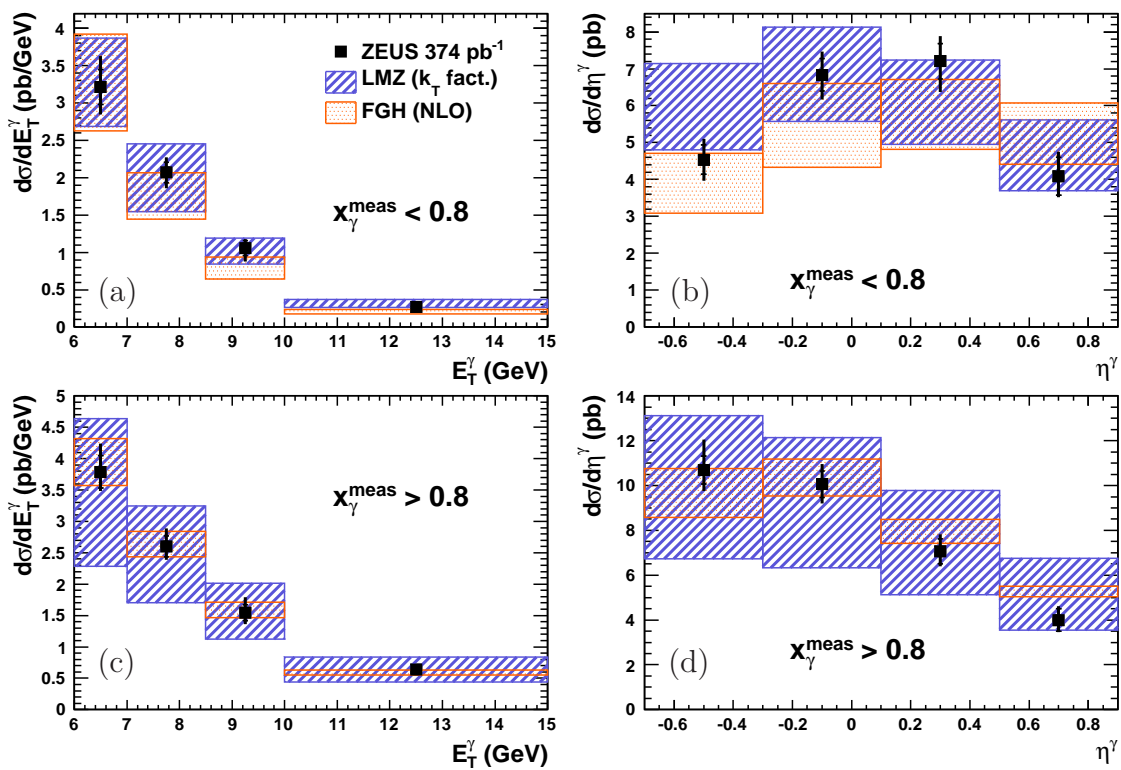,width=15cm}
\\[-68mm]
\hspace*{12mm}(a)\hspace*{70mm}(b)\hspace*{90mm}\\[43.5mm]
\hspace*{12mm}(c)\hspace*{70mm}(d)\hspace*{90mm}\\[16mm]

\caption{\small  Differential cross sections as functions of (a, c) \ETgam and (b, d) \etagam 
in different ranges of \xgamm, for events containing an isolated
photon accompanied by a jet, compared to predictions from FGH and LMZ.
The kinematic region of the measurement is described in the text.  The
inner and outer error bars respectively denote statistical
uncertainties and statistical uncertainties combined with systematic
uncertainties in quadrature. The theoretical uncertainties are shown
as hatched and dotted bands.  }
\label{fig:jetgam}
\end{figure}

\begin{figure}
\vfill
\epsfig{file=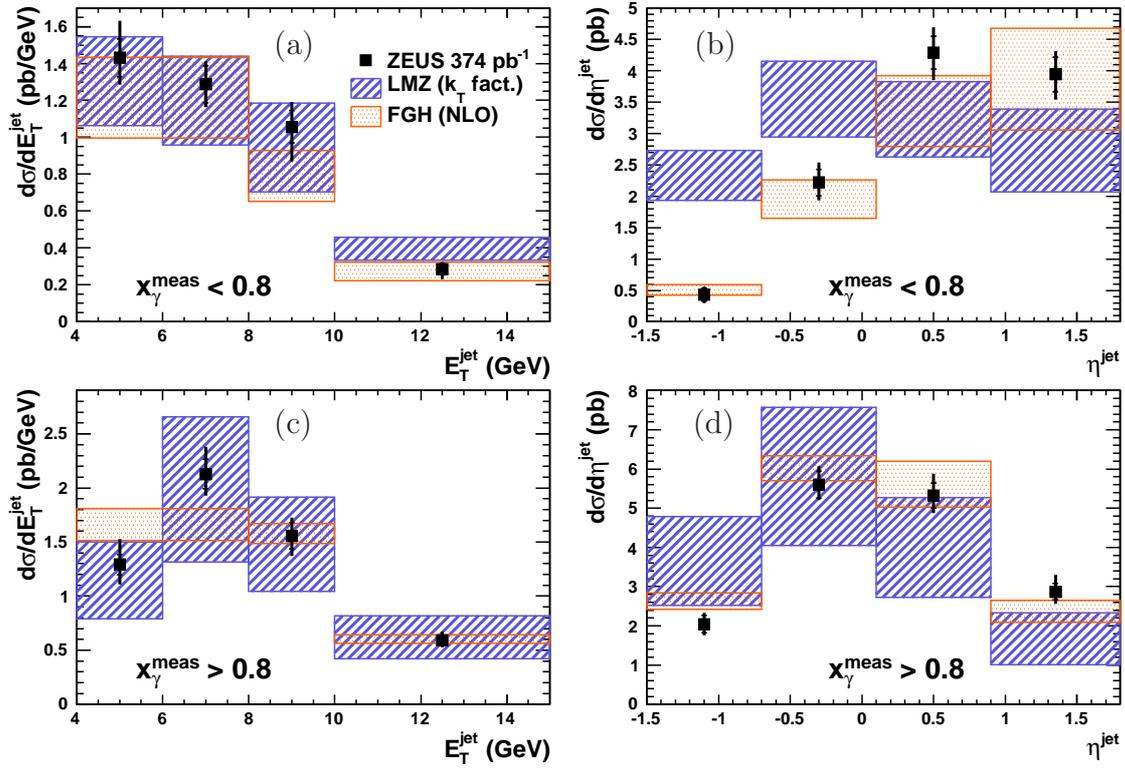,width=15cm}
\\[-100mm]
\hspace*{35mm}(a)\hspace*{50mm}(b)\hspace*{90mm}\\[44.5mm]
\hspace*{35mm}(c)\hspace*{50mm}(d)\hspace*{90mm}\\[46mm]

\caption{\small Differential cross sections as functions of (a, c) \ETjet and 
(b, d) \etajet, 
for events containing an isolated photon accompanied by a jet,
compared to predictions from FGH and LMZ.  
The first two FGH points in (a, c) have been averaged into
a single bin for calculational reasons.  
Other details as for Fig.~\ref{fig:jetgam}.}
\label{fig:jet}
\end{figure}

Complementing the previously published cross
sections~\cite{pl:b730:293} for the entire
\xgamm\ range, differential cross sections as functions of $E_T^{\gamma}$,
\etagam, \ETjet\ and \etajet\ are shown in
Figs.~\ref{fig:jetgam} -- \ref{fig:jet}. Here the selections of $\xgamm
> 0.8$ and $\xgamm<0.8$ have been applied to define ranges that
enhance the direct and resolved processes. In the \PYTHIA\ model
that was used, the upper and lower \xgamm\ ranges contain
direct and resolved events in the ratios 86:14 and 22:78,
respectively.

To within the theoretical uncertainties, the cross section predicted
by FGH is in quantitative agreement with the data; the LMZ predicted
cross section also agrees well for the photon and \ETjet\ variables,
but it is in disagreement with the \etajet\ distribution for $\xgamm <
0.8.$ This disagreement may be due to the modelling of the jet from
the parton cascade in the present version of the LMZ model.

\begin{figure}
\vspace*{-30mm}
\begin{center}
\epsfig{file=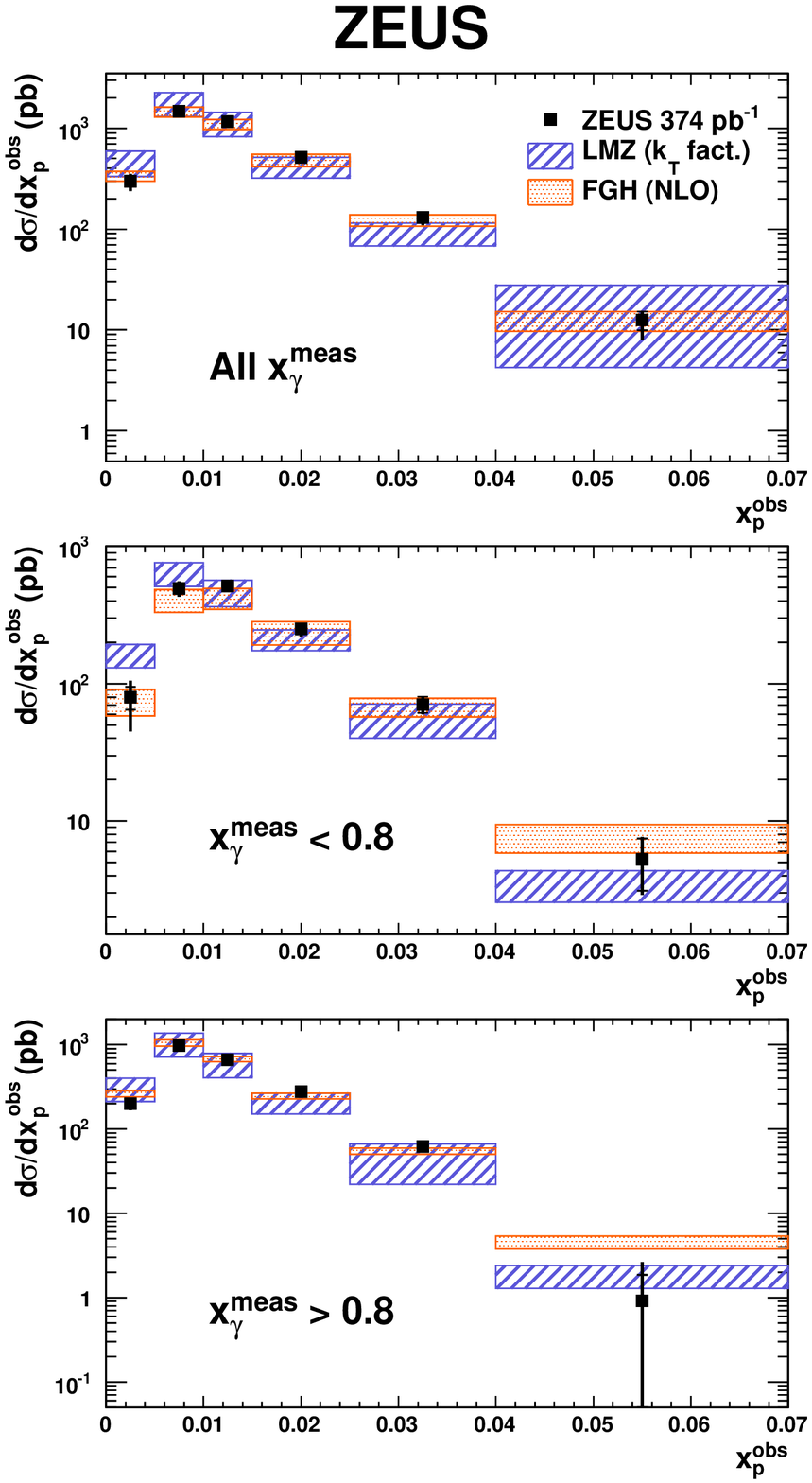,width=12cm}
\end{center}
~\\[-206mm]
\hspace*{8cm}(a)\\[61.5mm]
\hspace*{8cm}(b)\\[61.5mm]
\hspace*{8cm}(c)\\[61mm]
\caption{\small  Differential cross sections 
as functions of \xp\ for (a) all \xgamm, (b) $\xgamm>0.8$ (c)
$\xgamm<0.8$ for events containing an isolated photon
accompanied by a jet, compared to predictions from FGH and
LMZ.  Other details as for Fig.~\ref{fig:jetgam}.}
\label{fig:xp}
\end{figure}

\begin{figure}
\vspace*{-30mm}
\begin{center}
\epsfig{file=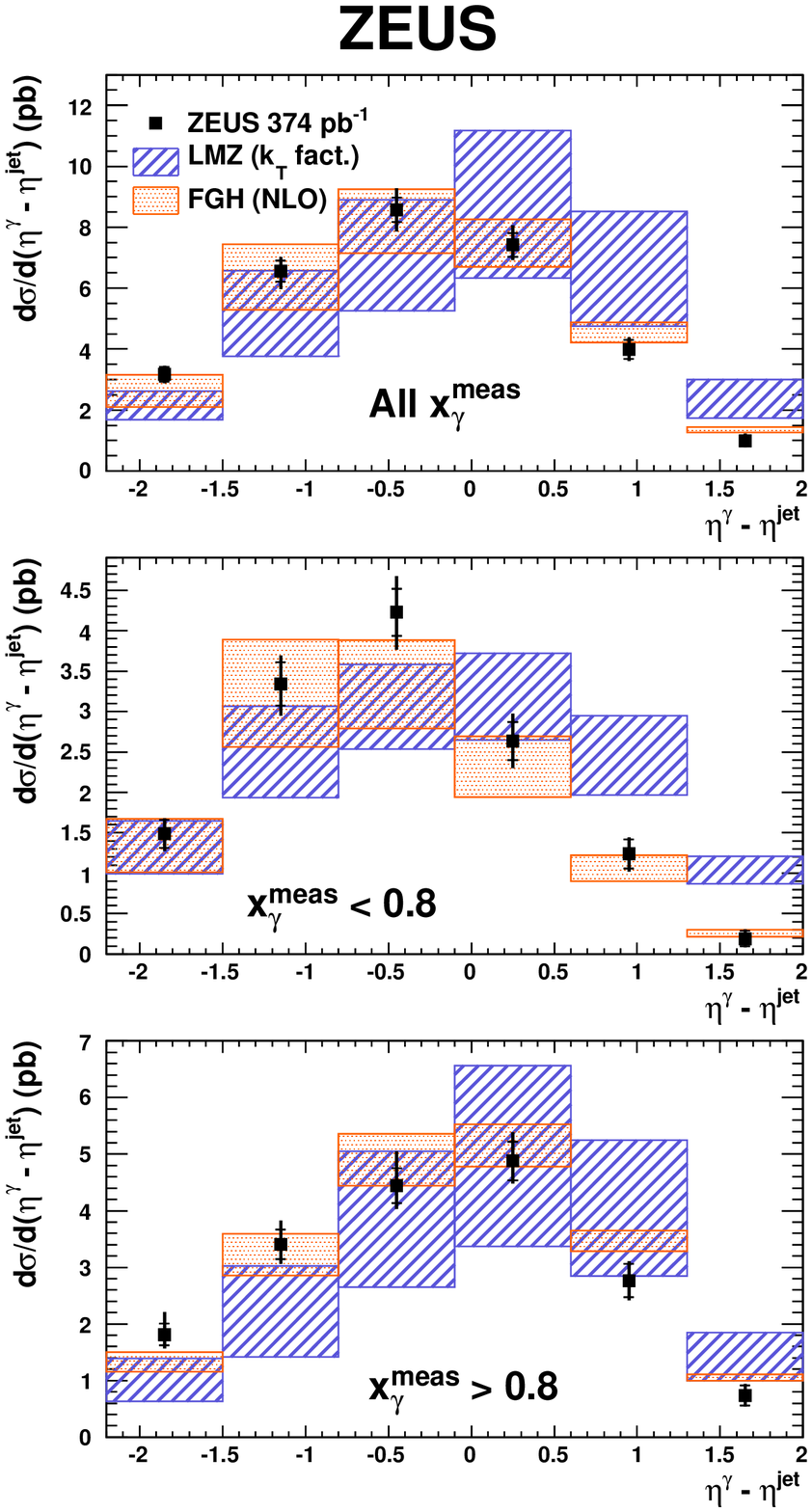,width=12cm}
\end{center}
~\\[-206mm]
\hspace*{12cm}(a)\\[61.5mm]
\hspace*{12cm}(b)\\[61.5mm]
\hspace*{12cm}(c)\\[61mm]
\caption{\small  Differential cross sections 
as functions of \deleta\ for (a) all \xgamm, (b) $\xgamm>0.8$ (c)
$\xgamm<0.8$ for events containing an isolated photon
accompanied by a jet, compared to predictions from FGH and
LMZ.  Other details as for Fig.~\ref{fig:jetgam}.}
\label{fig:deleta}
\end{figure}

The variables \xp\ and \deleta, presented in Figs.~\ref{fig:xp} and
\ref{fig:deleta}, also include results evaluated for the entire \xgamm\ range.
They are well described by FGH but slightly less so by LMZ.

\begin{figure}
\vspace*{-30mm}
\begin{center}
\epsfig{file=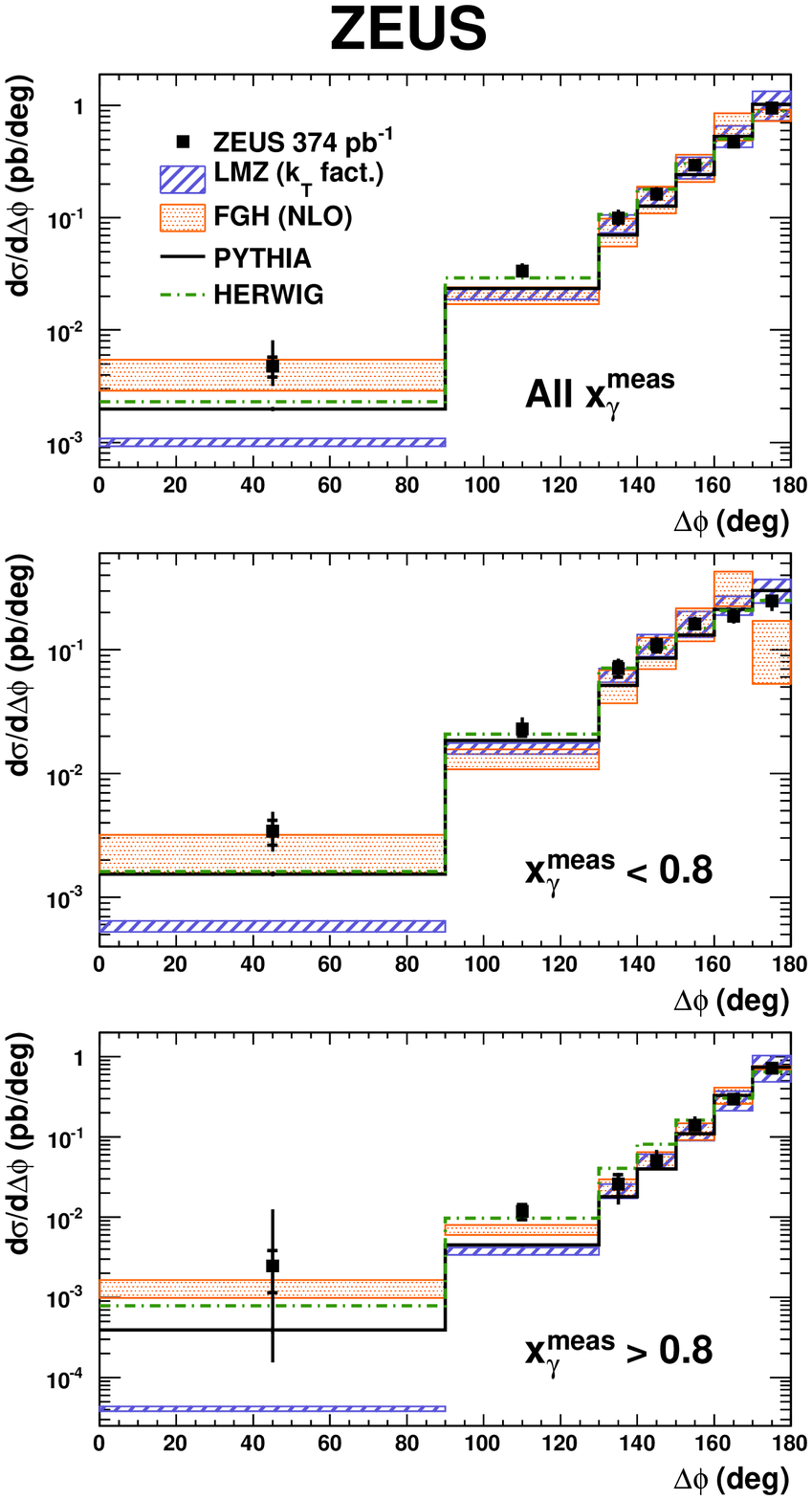,width=12cm}
\end{center}
~\\[-206mm]
\hspace*{8cm}(a)\\[61.5mm]
\hspace*{8cm}(b)\\[61.5mm]
\hspace*{8cm}(c)\\[61mm]
\caption{\small  Differential cross sections 
as functions of \delphi for (a) all \xgamm, (b) $\xgamm>0.8$ (c)
$\xgamm<0.8$ for events containing an isolated photon
accompanied by a jet, compared to predictions from FGH, LMZ, \PYTHIA\ and
\HERWIG.  Other details as for Fig.~\ref{fig:jetgam}.}
\label{fig:delphia}
\end{figure}

Differential cross sections as functions of \delphi\ are shown in
Fig.~\ref{fig:delphia}.  The data are compared to FGH and LMZ, with
similar conclusions as before, and are also compared to the versions
of \PYTHIA\ and \HERWIG\ described in Section \ref{sec-mc}.  The MC
programs both give a reasonable description of the data.  These
results demonstrate that parton showers used in conjunction with LO MC
programs can give a good description of higher-order contributions, as
also observed in other reactions \cite{mw}.

Tables \ref{tab:etg} to \ref{tab:deleta} give the numerical values
of the above results, together with the hadronisation factors that
were applied to the theory.  For further information, cross sections
calculated in the range $\xgamm<0.7$ are also listed.  These have a
direct:resolved ratio of 15:85 as modelled by \PYTHIA\ and show
features that are similar to the plotted results.

\section{Conclusions}

The production of  isolated photons with an
accompanying jet has been measured in photoproduction with the ZEUS
detector at HERA using an integrated luminosity of
$374\,\pm7\,\mathrm{pb}^{-1}$.  The present measurements extend earlier
ZEUS results.
Differential cross sections are presented in a kinematic region defined in the
laboratory frame by:\mbox{ $Q^2< 1 \,$}
$\mathrm{\Gev}^2$, $0.2 < y < 0.7$,
$-0.7< \eta^\gamma < 0.9$, $6 < E_T^{\gamma}< 15\,\mathrm{\Gev}$, $4< \ETjet <35$\gev\ and $-1.5 <\etajet<
1.8$.  Photon isolation was imposed such that at least $90 \%$ of the
energy of the jet-like object containing the photon originated from
the photon.

Differential cross sections are given in terms of the transverse
energy and pseudorapidity of the photon and the jet, and in terms of
\xp, \deleta and \delphi\ in high and low regions of \xgamm.  The
latter three variables are also presented for the entire observed
\xgamm\ range. The NLO-based predictions 
of Fontannaz, Guillet and Heinrich reproduce all the measured experimental
distributions well.  The $k_T$-factorisation approach of Lipatov,
Malyshev and Zotov describes most of the distributions well, with the
exception of the jet pseudorapidity at low \xgamm.

\section*{Acknowledgements}
\label{sec-ack}

\Zacknowledge\ We also thank M. Fontannaz, G. Heinrich, A. Lipatov, M. Malyshev and N. Zotov for providing assistance and theoretical results.

\vfill\eject

\raggedright{
\providecommand{\etal}{et al.\xspace}
\providecommand{\coll}{Collaboration}
\catcode`\@=11
\def\@bibitem#1{%
\ifmc@bstsupport
  \mc@iftail{#1}%
    {;\newline\ignorespaces}%
    {\ifmc@first\else.\fi\orig@bibitem{#1}}
  \mc@firstfalse
\else
  \mc@iftail{#1}%
    {\ignorespaces}%
    {\orig@bibitem{#1}}%
\fi}%
\catcode`\@=12

}

\begin{table}
\begin{center}
\begin{tabular}{|rcr|c|c|}
\hline
\multicolumn{3}{|c|}{$E_T^{\gamma}$ range}   &       & \\[-1.8ex]
\multicolumn{3}{|c|}{ (GeV)}   &  
 \raisebox{1.8ex}{$\frac{d\sigma}{dE^{\gamma}_T}$ ($\mathrm{pb}\,\mathrm{GeV}^{-1})$} &    
\raisebox{1.8ex}{had.\ corr.\!} \\
\hline\hline
\multicolumn{3}{|l|}{$\xgamm>0.8$}   & & \\
\hline
6.0 & -- & 7.0   & $3.79 \pm  0.26\,\mathrm{(stat.)}\,^{+0.37}_{-0.16}\,\mathrm{(syst.)}$ &0.84 \\
7.0 & -- & 8.5   & $2.60 \pm  0.17\,\mathrm{(stat.)}\,^{+0.23}_{-0.14}\,\mathrm{(syst.)}$ &0.90 \\
8.5 & -- & 10.0   & $1.55 \pm  0.13\,\mathrm{(stat.)}\,^{+0.21}_{-0.13}\,\mathrm{(syst.)}$ &0.96 \\
10.0 & -- & 15.0   & $0.63 \pm  0.04\,\mathrm{(stat.)}\,^{+0.04}_{-0.04}\,\mathrm{(syst.)}$ &0.98 \\

\hline
\multicolumn{3}{|l|}{$\xgamm<0.8$}   & & \\
\hline
6.0 & -- & 7.0   & $3.22 \pm  0.24\,\mathrm{(stat.)}\,^{+0.34}_{-0.28}\,\mathrm{(syst.)}$ &0.79 \\
7.0 & -- & 8.5   & $2.07 \pm  0.14\,\mathrm{(stat.)}\,^{+0.15}_{-0.16}\,\mathrm{(syst.)}$ &0.80 \\
8.5 & -- & 10.0   & $1.06 \pm  0.10\,\mathrm{(stat.)}\,^{+0.07}_{-0.15}\,\mathrm{(syst.)}$ &0.81 \\
10.0 & -- & 15.0   & $0.27 \pm  0.03\,\mathrm{(stat.)}\,^{+0.02}_{-0.03}\,\mathrm{(syst.)}$ &0.83 \\

\hline
\multicolumn{3}{|l|}{$\xgamm<0.7$}   & & \\
\hline
6.0 & -- & 7.0   & $2.37 \pm  0.21\,\mathrm{(stat.)}\,^{+0.29}_{-0.21}\,\mathrm{(syst.)}$ &0.72 \\
7.0 & -- & 8.5   & $1.32 \pm  0.12\,\mathrm{(stat.)}\,^{+0.10}_{-0.09}\,\mathrm{(syst.)}$ &0.75 \\
8.5 & -- & 10.0   & $0.66 \pm  0.09\,\mathrm{(stat.)}\,^{+0.07}_{-0.08}\,\mathrm{(syst.)}$ &0.77 \\
10.0 & -- & 15.0   & $0.18 \pm  0.03\,\mathrm{(stat.)}\,^{+0.02}_{-0.03}\,\mathrm{(syst.)}$ &0.80 \\

\hline

\end{tabular}
\end{center}
\caption{Differential cross-section $\frac{d\sigma}{dE^{\gamma}_T}$ for photons accompanied 
by a jet, and hadronisation correction. \label{tab:etg}}
\end{table}

\begin{table}
\begin{center}
\begin{tabular}{|rcr|c|c|}
\hline
\multicolumn{3}{|c|}{ $\eta^{\gamma}$ range }  &  $\frac{d\sigma}{d\eta^{\gamma}}$ ($\mathrm{pb}$)     & had.\ corr.\! \\[0.5mm]
\hline\hline
\multicolumn{3}{|l|}{$\xgamm>0.8$}   & & \\
\hline
--\,0.7 & -- & --\,0.3   & $10.69 \pm  0.62\,\mathrm{(stat.)}\,^{+1.20}_{-0.71}\,\mathrm{(syst.)}$ &0.93 \\
--\,0.3 & -- & 0.1   & $10.07 \pm  0.59\,\mathrm{(stat.)}\,^{+0.66}_{-0.63}\,\mathrm{(syst.)}$ &0.93 \\
0.1 & -- & 0.5   & $7.06 \pm  0.56\,\mathrm{(stat.)}\,^{+0.51}_{-0.36}\,\mathrm{(syst.)}$ &0.90 \\
0.5 & -- & 0.9   & $4.00 \pm  0.50\,\mathrm{(stat.)}\,^{+0.36}_{-0.20}\,\mathrm{(syst.)}$ &0.87 \\
\hline
\multicolumn{3}{|l|}{$\xgamm<0.8$}   & & \\
\hline
--\,0.7 & -- & --\,0.3   & $4.54 \pm  0.40\,\mathrm{(stat.)}\,^{+0.41}_{-0.42}\,\mathrm{(syst.)}$ &0.84 \\
--\,0.3 & -- & 0.1   & $6.83 \pm  0.44\,\mathrm{(stat.)}\,^{+0.46}_{-0.49}\,\mathrm{(syst.)}$ &0.80 \\
0.1 & -- & 0.5   & $7.20 \pm  0.48\,\mathrm{(stat.)}\,^{+0.47}_{-0.68}\,\mathrm{(syst.)}$ &0.80 \\
0.5 & -- & 0.9   & $4.08 \pm  0.51\,\mathrm{(stat.)}\,^{+0.43}_{-0.21}\,\mathrm{(syst.)}$ &0.79 \\

\hline
\multicolumn{3}{|l|}{$\xgamm<0.7$}   & & \\
\hline
--\,0.7 & -- & --\,0.3   & $2.79 \pm  0.31\,\mathrm{(stat.)}\,^{+0.26}_{-0.27}\,\mathrm{(syst.)}$ &0.78 \\
--\,0.3 & -- & 0.1   & $4.56 \pm  0.38\,\mathrm{(stat.)}\,^{+0.31}_{-0.31}\,\mathrm{(syst.)}$ &0.74 \\
0.1 & -- & 0.5   & $5.12 \pm  0.44\,\mathrm{(stat.)}\,^{+0.32}_{-0.52}\,\mathrm{(syst.)}$ &0.74 \\
0.5 & -- & 0.9   & $3.15 \pm  0.49\,\mathrm{(stat.)}\,^{+0.43}_{-0.26}\,\mathrm{(syst.)}$ &0.74 \\
\hline

\end{tabular}
\end{center}
\caption{Differential cross-section $\frac{d\sigma}{d\eta^{\gamma}}$ for photons accompanied by a jet, and hadronisation correction.\label{tab:etag}}
\end{table}

\begin{table}
\begin{center}
\begin{tabular}{|rcr|c|c|}
\hline
\multicolumn{3}{|c|}{$E_T^{\jet}$ range}   &       & \\[-1.8ex]
\multicolumn{3}{|c|}{ (GeV)}   &  \raisebox{1.8ex}{$\frac{d\sigma}{dE^{\jet}_T}$ ($\mathrm{pb}\,\mathrm{GeV}^{-1})$}   &  
\raisebox{1.8ex}{had.\ corr.\!} \\ 
\hline\hline
\multicolumn{3}{|l|}{$\xgamm>0.8$}   & & \\
\hline
4.0 & -- & 6.0   & $1.29 \pm  0.10\,\mathrm{(stat.)}\,^{+0.21}_{-0.16}\,\mathrm{(syst.)}$ &0.81 \\
6.0 & -- & 8.0   & $2.13 \pm  0.14\,\mathrm{(stat.)}\,^{+0.21}_{-0.15}\,\mathrm{(syst.)}$ &0.83 \\
8.0 & -- & 10.0   & $1.56 \pm  0.12\,\mathrm{(stat.)}\,^{+0.12}_{-0.14}\,\mathrm{(syst.)}$ &0.96 \\
10.0 & -- & 15.0   & $0.59 \pm  0.04\,\mathrm{(stat.)}\,^{+0.07}_{-0.05}\,\mathrm{(syst.)}$ &1.05 \\
\hline
\multicolumn{3}{|l|}{$\xgamm<0.8$}   & & \\
\hline
4.0 & -- & 6.0   & $1.43 \pm  0.10\,\mathrm{(stat.)}\,^{+0.17}_{-0.10}\,\mathrm{(syst.)}$ &0.84 \\
6.0 & -- & 8.0   & $1.29 \pm  0.10\,\mathrm{(stat.)}\,^{+0.08}_{-0.07}\,\mathrm{(syst.)}$ &0.73 \\
8.0 & -- & 10.0   & $1.06 \pm  0.09\,\mathrm{(stat.)}\,^{+0.10}_{-0.17}\,\mathrm{(syst.)}$ &0.80 \\
10.0 & -- & 15.0   & $0.28 \pm  0.03\,\mathrm{(stat.)}\,^{+0.02}_{-0.04}\,\mathrm{(syst.)}$ &0.87 \\
\hline
\multicolumn{3}{|l|}{$\xgamm<0.7$}   & & \\
\hline
4.0 & -- & 6.0   & $1.07 \pm  0.09\,\mathrm{(stat.)}\,^{+0.15}_{-0.08}\,\mathrm{(syst.)}$ &0.76 \\
6.0 & -- & 8.0   & $0.82 \pm  0.09\,\mathrm{(stat.)}\,^{+0.05}_{-0.05}\,\mathrm{(syst.)}$ &0.68 \\
8.0 & -- & 10.0   & $0.73 \pm  0.08\,\mathrm{(stat.)}\,^{+0.07}_{-0.14}\,\mathrm{(syst.)}$ &0.77 \\
10.0 & -- & 15.0   & $0.20 \pm  0.03\,\mathrm{(stat.)}\,^{+0.02}_{-0.03}\,\mathrm{(syst.)}$ &0.83 \\

\hline
\end{tabular}
\end{center}
\caption{Differential cross-section
$\frac{d\sigma}{dE^{\jet}_T}$ for photons accompanied by a jet, and
hadronisation correction.
\label{tab:jetgam}}
\end{table}

\begin{table}
\begin{center}
\begin{tabular}{|rcr|c|c|}
\hline
\multicolumn{3}{|c|}{ $\eta^{\jet}$ range } &\multicolumn{1}{|c|}{ $\frac{d\sigma}{d\eta^{\jet}}$ 
($\mathrm{pb}$)}     & had.\ corr.\! \\[0.5mm]
\hline\hline
\multicolumn{3}{|l|}{$\xgamm>0.8$}   & & \\
\hline
--1.5 & -- & --\,0.7   & $2.04 \pm  0.22\,\mathrm{(stat.)}\,^{+0.18}_{-0.18}\,\mathrm{(syst.)}$ &0.68 \\
--\,0.7 & -- & 0.1   & $5.60 \pm  0.35\,\mathrm{(stat.)}\,^{+0.31}_{-0.18}\,\mathrm{(syst.)}$ &0.83 \\
0.1 & -- & 0.9   & $5.32 \pm  0.32\,\mathrm{(stat.)}\,^{+0.45}_{-0.32}\,\mathrm{(syst.)}$ &1.09 \\
0.9 & -- & 1.8   & $2.87 \pm  0.21\,\mathrm{(stat.)}\,^{+0.38}_{-0.23}\,\mathrm{(syst.)}$ &1.33 \\

\hline
\multicolumn{3}{|l|}{$\xgamm<0.8$}   & & \\
\hline

--1.5 & -- & --\,0.7   & $0.43 \pm  0.10\,\mathrm{(stat.)}\,^{+0.07}_{-0.09}\,\mathrm{(syst.)}$ &1.15 \\
--\,0.7 & -- & 0.1   & $2.22 \pm  0.21\,\mathrm{(stat.)}\,^{+0.25}_{-0.19}\,\mathrm{(syst.)}$ &0.79 \\
0.1 & -- & 0.9   & $4.29 \pm  0.26\,\mathrm{(stat.)}\,^{+0.31}_{-0.35}\,\mathrm{(syst.)}$ &0.73 \\
0.9 & -- & 1.8   & $3.94 \pm  0.27\,\mathrm{(stat.)}\,^{+0.24}_{-0.30}\,\mathrm{(syst.)}$ &0.85 \\
\hline
\multicolumn{3}{|l|}{$\xgamm<0.7$}   & & \\
\hline

--1.5 & -- & --\,0.7   & $0.08 \pm  0.08\,\mathrm{(stat.)}\,^{+0.08}_{-0.05}\,\mathrm{(syst.)}$ &0.83 \\
--\,0.7 & -- & 0.1   & $1.18 \pm  0.17\,\mathrm{(stat.)}\,^{+0.14}_{-0.08}\,\mathrm{(syst.)}$ &0.69 \\
0.1 & -- & 0.9   & $3.11 \pm  0.23\,\mathrm{(stat.)}\,^{+0.22}_{-0.26}\,\mathrm{(syst.)}$ &0.69 \\
0.9 & -- & 1.8   & $3.05 \pm  0.25\,\mathrm{(stat.)}\,^{+0.22}_{-0.24}\,\mathrm{(syst.)}$ &0.82 \\

\hline
\end{tabular}
\end{center}
\caption{Differential cross-section
$\frac{d\sigma}{d\eta^{\jet}}$ for photons accompanied by a jet, and
hadronisation correction.
\label{tab:etaj}}
\end{table}

\begin{table}
\begin{center}
\begin{tabular}{|rcr|c|c|}
\hline
\multicolumn{3}{|c|}{ $\xp$ range } &   $\frac{d\sigma}{d\xp}$ ($\mathrm{pb}$)
& had.\ corr.\! \\[1.0mm]
\hline\hline
\multicolumn{3}{|l|}{All $\xgamm$}   & & \\
\hline
0.0 & -- & 0.005   &  $297.6 \pm  30.4\,\mathrm{(stat.)}\,^{+46.0}_{-51.3}\,\mathrm{(syst.)}$ &0.76 \\
0.005 & -- & 0.010   & $1471.5 \pm  63.2\,\mathrm{(stat.)}\,^{+135.3}_{-124.3}\,\mathrm{(syst.)}$ &0.80 \\
0.010 & -- & 0.015   & $1160.0 \pm  57.5\,\mathrm{(stat.)}\,^{+56.9}_{-57.2}\,\mathrm{(syst.)}$ &0.90 \\
0.015 & -- & 0.025   & $514.5 \pm  27.8\,\mathrm{(stat.)}\,^{+20.5}_{-29.4}\,\mathrm{(syst.)}$ &0.94 \\
0.025 & -- & 0.040   & $130.1 \pm  11.7\,\mathrm{(stat.)}\,^{+6.6}_{-17.1}\,\mathrm{(syst.)}$ &0.99 \\
0.040 & -- & 0.070   & $12.6 \pm  2.6\,\mathrm{(stat.)}\,^{+1.0}_{-3.8}\,\mathrm{(syst.)}$ &1.00 \\
\hline
\multicolumn{3}{|l|}{$\xgamm>0.8$}   & & \\
\hline
0.0 & -- & 0.005   & $199.5 \pm  27.3\,\mathrm{(stat.)}\,^{+23.2}_{-15.3}\,\mathrm{(syst.)}$ &0.72 \\
0.005 & -- & 0.010   & $975.3 \pm  54.4\,\mathrm{(stat.)}\,^{+81.8}_{-68.1}\,\mathrm{(syst.)}$ &0.82 \\
0.010 & -- & 0.015   & $662.8 \pm  46.7\,\mathrm{(stat.)}\,^{+68.4}_{-28.8}\,\mathrm{(syst.)}$ &1.00 \\
0.015 & -- & 0.025   & $276.9 \pm  21.3\,\mathrm{(stat.)}\,^{+21.4}_{-15.3}\,\mathrm{(syst.)}$ &1.12 \\
0.025 & -- & 0.040   & $61.9 \pm  8.0\,\mathrm{(stat.)}\,^{+3.6}_{-4.3}\,\mathrm{(syst.)}$ &1.26 \\
0.040 & -- & 0.070   & $0.9 \pm  0.9\,\mathrm{(stat.)}\,^{+1.5}_{-0.6}\,\mathrm{(syst.)}$ &1.29 \\
\hline
\multicolumn{3}{|l|}{$\xgamm<0.8$}   & & \\
\hline
0.0 & -- & 0.005   & $79.6 \pm  14.8\,\mathrm{(stat.)}\,^{+20.7}_{-31.4}\,\mathrm{(syst.)}$ &0.95 \\
0.005 & -- & 0.010   & $492.3 \pm  37.0\,\mathrm{(stat.)}\,^{+52.3}_{-53.5}\,\mathrm{(syst.)}$ &0.77 \\
0.010 & -- & 0.015   & $515.2 \pm  38.1\,\mathrm{(stat.)}\,^{+24.7}_{-28.0}\,\mathrm{(syst.)}$ &0.78 \\
0.015 & -- & 0.025   & $249.9 \pm  20.6\,\mathrm{(stat.)}\,^{+14.0}_{-21.9}\,\mathrm{(syst.)}$ &0.81 \\
0.025 & -- & 0.040   & $70.9 \pm  9.4\,\mathrm{(stat.)}\,^{+3.7}_{-6.1}\,\mathrm{(syst.)}$ &0.85 \\
0.040 & -- & 0.070   & $5.3 \pm  2.2\,\mathrm{(stat.)}\,^{+0.9}_{-1.0}\,\mathrm{(syst.)}$ &0.86 \\
\hline
\multicolumn{3}{|l|}{$\xgamm<0.7$}   & & \\
\hline
0.0 & -- & 0.005   & $35.5 \pm  11.4\,\mathrm{(stat.)}\,^{+9.4}_{-10.4}\,\mathrm{(syst.)}$ &0.69 \\
0.005 & -- & 0.010   & $298.3 \pm  30.4\,\mathrm{(stat.)}\,^{+34.9}_{-39.5}\,\mathrm{(syst.)}$ &0.68 \\
0.010 & -- & 0.015   & $366.1 \pm  33.5\,\mathrm{(stat.)}\,^{+21.0}_{-21.6}\,\mathrm{(syst.)}$ &0.73 \\
0.015 & -- & 0.025   & $193.6 \pm  18.8\,\mathrm{(stat.)}\,^{+12.6}_{-21.6}\,\mathrm{(syst.)}$ &0.78 \\
0.025 & -- & 0.040   & $51.4 \pm  9.0\,\mathrm{(stat.)}\,^{+2.1}_{-4.8}\,\mathrm{(syst.)}$ &0.83 \\
0.040 & -- & 0.070   & $3.8 \pm  2.1\,\mathrm{(stat.)}\,^{+1.1}_{-0.9}\,\mathrm{(syst.)}$ &0.82 \\
\hline
\end{tabular}
\end{center}
\caption{Differential cross-section $\frac{d\sigma}{d\xp}$
for photons accompanied by a jet, and hadronisation correction.
\label{tab:xp}}
\end{table}

\begin{table}
\begin{center}
\begin{tabular}{|rrr|c|c|}
\hline
\multicolumn{3}{|c|}{$(\eta^{\gamma}-\eta^\jet) $  range }& 
 $\frac{d\sigma}{d(\eta^{\gamma}-\eta^\jet)}$ ($\mathrm{pb}$)  & had.\ corr.\!\\[0.5mm]
\hline\hline
\multicolumn{3}{|l|}{All $\xgamm$}   & & \\
\hline
\hspace*{1.0ex}--2.2 & \hspace*{1.0ex}-- &--1.5   & $3.17 \pm  0.24\,\mathrm{(stat.)}\,^{+0.14}_{-0.18}\,\mathrm{(syst.)}$ &1.04 \\
--1.5 & -- &--\,0.8   & $6.56 \pm  0.35\,\mathrm{(stat.)}\,^{+0.31}_{-0.48}\,\mathrm{(syst.)}$ &0.96 \\
--\,0.8 & -- &--\,0.1   & $8.57 \pm  0.40\,\mathrm{(stat.)}\,^{+0.58}_{-0.59}\,\mathrm{(syst.)}$ &0.89 \\
--\,0.1 & -- & 0.6   & $7.42 \pm  0.38\,\mathrm{(stat.)}\,^{+0.52}_{-0.31}\,\mathrm{(syst.)}$ &0.84 \\
0.6 & -- & 1.3   & $3.99 \pm  0.32\,\mathrm{(stat.)}\,^{+0.23}_{-0.22}\,\mathrm{(syst.)}$ &0.77 \\
1.3 & -- & 2.0   & $0.98 \pm  0.19\,\mathrm{(stat.)}\,^{+0.14}_{-0.07}\,\mathrm{(syst.)}$ &0.73 \\
\hline
\multicolumn{3}{|l|}{$\xgamm>0.8$}   & & \\
\hline
--2.2 & -- & --1.5   & $1.81 \pm  0.19\,\mathrm{(stat.)}\,^{+0.35}_{-0.15}\,\mathrm{(syst.)}$ &1.32 \\
--1.5 & -- & --\,0.8   & $3.41 \pm  0.26\,\mathrm{(stat.)}\,^{+0.33}_{-0.23}\,\mathrm{(syst.)}$ &1.18 \\
--\,0.8 & -- & --\,0.1   & $4.44 \pm  0.31\,\mathrm{(stat.)}\,^{+0.53}_{-0.27}\,\mathrm{(syst.)}$ &1.04 \\
--\,0.1 & -- & 0.6   & $4.88 \pm  0.34\,\mathrm{(stat.)}\,^{+0.37}_{-0.21}\,\mathrm{(syst.)}$ &0.88 \\
0.6 & -- & 1.3   & $2.77 \pm  0.29\,\mathrm{(stat.)}\,^{+0.18}_{-0.18}\,\mathrm{(syst.)}$ &0.74 \\
1.3 & -- & 2.0   & $0.74 \pm  0.18\,\mathrm{(stat.)}\,^{+0.09}_{-0.09}\,\mathrm{(syst.)}$ &0.65 \\
\hline
\multicolumn{3}{|l|}{$\xgamm<0.8$}   & & \\
\hline
--2.2 & -- & --1.5   & $1.49 \pm  0.17\,\mathrm{(stat.)}\,^{+0.08}_{-0.12}\,\mathrm{(syst.)}$ &0.89 \\
--1.5 & -- & --\,0.8   & $3.34 \pm  0.27\,\mathrm{(stat.)}\,^{+0.23}_{-0.29}\,\mathrm{(syst.)}$ &0.83 \\
--\,0.8 & -- & --\,0.1   & $4.23 \pm  0.29\,\mathrm{(stat.)}\,^{+0.34}_{-0.36}\,\mathrm{(syst.)}$ &0.75 \\
--\,0.1 & -- & 0.6   & $2.63 \pm  0.24\,\mathrm{(stat.)}\,^{+0.24}_{-0.23}\,\mathrm{(syst.)}$ &0.76 \\
0.6 & -- & 1.3   & $1.24 \pm  0.18\,\mathrm{(stat.)}\,^{+0.10}_{-0.13}\,\mathrm{(syst.)}$ &0.88 \\
1.3 & -- & 2.0   & $0.19 \pm  0.09\,\mathrm{(stat.)}\,^{+0.07}_{-0.05}\,\mathrm{(syst.)}$ &1.15 \\
\hline
\multicolumn{3}{|l|}{$\xgamm<0.7$}   & & \\
\hline

--2.2 & -- & --1.5   & $1.02 \pm  0.15\,\mathrm{(stat.)}\,^{+0.08}_{-0.08}\,\mathrm{(syst.)}$ &0.85 \\
--1.5 & -- & --\,0.8   & $2.56 \pm  0.25\,\mathrm{(stat.)}\,^{+0.18}_{-0.25}\,\mathrm{(syst.)}$ &0.81 \\
--\,0.8 & -- & --\,0.1   & $3.19 \pm  0.27\,\mathrm{(stat.)}\,^{+0.24}_{-0.27}\,\mathrm{(syst.)}$ &0.72 \\
--\,0.1 & -- & 0.6   & $1.69 \pm  0.21\,\mathrm{(stat.)}\,^{+0.24}_{-0.11}\,\mathrm{(syst.)}$ &0.68 \\
0.6 & -- & 1.3   & $0.61 \pm  0.15\,\mathrm{(stat.)}\,^{+0.08}_{-0.10}\,\mathrm{(syst.)}$ &0.71 \\
1.3 & -- & 2.0   & $0.00 \pm  0.48\,\mathrm{(stat.)}\,^{+0.13}_{-0.00}\,\mathrm{(syst.)}$ &0.87 \\
\hline
\end{tabular}
\end{center}
\caption{Differential cross-section $\frac{d\sigma}{d(\eta^{\gamma}-\eta^\jet)}$ 
for photons accompanied by a jet, and hadronisation correction.
\label{tab:deleta}}
\end{table}

\begin{table}
\begin{center}
\begin{tabular}{|rcr|c|c|}
\hline
\multicolumn{3}{|c|}{ $\Delta\phi$ range } & & \\[-1.8ex]
\multicolumn{3}{|c|}{ (deg.)}   &  
\raisebox{1.8ex}{$\frac{d\sigma}{d\Delta\phi}$ ($\mathrm{pb}\,\mathrm{deg.}^{-1})$}   &  
\raisebox{1.8ex}{had.\ corr.\!}\\ 
\hline\hline
\multicolumn{3}{|l|}{All $\xgamm$}   & & \\
\hline
0.0 & -- & 90.0   & $0.0048 \pm  0.0010\,\mathrm{(stat.)}\,^{+0.0032}_{-0.0013}\,\mathrm{(syst.)}$ &0.78 \\
90.0 & -- & 130.0   & $0.033 \pm  0.004\,\mathrm{(stat.)}\,^{+0.005}_{-0.002}\,\mathrm{(syst.)}$ &0.81 \\
130.0 & -- & 140.0   & $0.100 \pm  0.012\,\mathrm{(stat.)}\,^{+0.013}_{-0.009}\,\mathrm{(syst.)}$ &0.82 \\
140.0 & -- & 150.0   & $0.164 \pm  0.016\,\mathrm{(stat.)}\,^{+0.018}_{-0.014}\,\mathrm{(syst.)}$ &0.85 \\
150.0 & -- & 160.0   & $0.296 \pm  0.019\,\mathrm{(stat.)}\,^{+0.027}_{-0.016}\,\mathrm{(syst.)}$ &0.86 \\
160.0 & -- & 170.0   & $0.473 \pm  0.026\,\mathrm{(stat.)}\,^{+0.019}_{-0.026}\,\mathrm{(syst.)}$ &0.89 \\
170.0 & -- & 180.0   & $0.951 \pm  0.036\,\mathrm{(stat.)}\,^{+0.030}_{-0.066}\,\mathrm{(syst.)}$ &0.86 \\
\hline
\multicolumn{3}{|l|}{$\xgamm>0.8$}   & & \\
\hline

0.0 & -- & 90.0   & $0.002 \pm  0.001\,\mathrm{(stat.)}\,^{+0.010}_{-0.002}\,\mathrm{(syst.)}$ &0.57 \\
90.0 & -- & 130.0   & $0.012 \pm  0.003\,\mathrm{(stat.)}\,^{+0.001}_{-0.001}\,\mathrm{(syst.)}$ &0.76 \\
130.0 & -- & 140.0   & $0.026 \pm  0.008\,\mathrm{(stat.)}\,^{+0.005}_{-0.009}\,\mathrm{(syst.)}$ &0.77 \\
140.0 & -- & 150.0   & $0.051 \pm  0.010\,\mathrm{(stat.)}\,^{+0.015}_{-0.006}\,\mathrm{(syst.)}$ &0.85 \\
150.0 & -- & 160.0   & $0.140 \pm  0.014\,\mathrm{(stat.)}\,^{+0.037}_{-0.006}\,\mathrm{(syst.)}$ &0.89 \\
160.0 & -- & 170.0   & $0.295 \pm  0.022\,\mathrm{(stat.)}\,^{+0.014}_{-0.033}\,\mathrm{(syst.)}$ &0.93 \\
170.0 & -- & 180.0   & $0.720 \pm  0.034\,\mathrm{(stat.)}\,^{+0.045}_{-0.055}\,\mathrm{(syst.)}$ &0.91 \\
\hline
\multicolumn{3}{|l|}{$\xgamm<0.8$}   & & \\
\hline
0.0 & -- & 90.0   & $0.0034 \pm  0.0008\,\mathrm{(stat.)}\,^{+0.0013}_{-0.0007}\,\mathrm{(syst.)}$ &0.79 \\
90.0 & -- & 130.0   & $0.0230 \pm  0.0030\,\mathrm{(stat.)}\,^{+0.0045}_{-0.0014}\,\mathrm{(syst.)}$ &0.82 \\
130.0 & -- & 140.0   & $0.070 \pm  0.010\,\mathrm{(stat.)}\,^{+0.011}_{-0.007}\,\mathrm{(syst.)}$ &0.84 \\
140.0 & -- & 150.0   & $0.110 \pm  0.014\,\mathrm{(stat.)}\,^{+0.009}_{-0.008}\,\mathrm{(syst.)}$ &0.86 \\
150.0 & -- & 160.0   & $0.162 \pm  0.015\,\mathrm{(stat.)}\,^{+0.018}_{-0.009}\,\mathrm{(syst.)}$ &0.84 \\
160.0 & -- & 170.0   & $0.187 \pm  0.017\,\mathrm{(stat.)}\,^{+0.011}_{-0.017}\,\mathrm{(syst.)}$ &0.82 \\
170.0 & -- & 180.0   & $0.247 \pm  0.020\,\mathrm{(stat.)}\,^{+0.016}_{-0.035}\,\mathrm{(syst.)}$ &0.76 \\
\hline
\multicolumn{3}{|l|}{$\xgamm<0.7$}   & & \\
\hline
0.0 & -- & 90.0   & $0.0023 \pm  0.0006\,\mathrm{(stat.)}\,^{+0.0010}_{-0.0005}\,\mathrm{(syst.)}$ &0.75 \\
90.0 & -- & 130.0   & $0.0168 \pm  0.0027\,\mathrm{(stat.)}\,^{+0.0051}_{-0.0015}\,\mathrm{(syst.)}$ &0.78 \\
130.0 & -- & 140.0   & $0.046 \pm  0.008\,\mathrm{(stat.)}\,^{+0.006}_{-0.004}\,\mathrm{(syst.)}$ &0.80 \\
140.0 & -- & 150.0   & $0.063 \pm  0.012\,\mathrm{(stat.)}\,^{+0.016}_{-0.005}\,\mathrm{(syst.)}$ &0.79 \\
150.0 & -- & 160.0   & $0.104 \pm  0.013\,\mathrm{(stat.)}\,^{+0.007}_{-0.007}\,\mathrm{(syst.)}$ &0.77 \\
160.0 & -- & 170.0   & $0.133 \pm  0.015\,\mathrm{(stat.)}\,^{+0.008}_{-0.012}\,\mathrm{(syst.)}$ &0.76 \\
170.0 & -- & 180.0   & $0.172 \pm  0.017\,\mathrm{(stat.)}\,^{+0.010}_{-0.026}\,\mathrm{(syst.)}$ &0.70 \\
\hline
\end{tabular}
\end{center}
\caption{Differential cross-section $\frac{d\sigma}{d\Delta\phi}$
for photons accompanied by a jet, and hadronisation correction.
\label{tab:dphi}}
\end{table}

\newpage
\clearpage
%

%

%

\begin{mcbibliography}{10}
\bibitem{pl:b730:293}
ZEUS \coll, H. Abramowicz \etal,
 Phys.\ Lett.\ B~730~(2014)~293\relax 
\relax
\bibitem{pl:b413:201}
ZEUS \coll, J.~Breitweg \etal,
\newblock Phys.\ Lett.{} B~413~(1997)~201\relax
\relax
\bibitem{pl:b472:175}
ZEUS \coll, J.~Breitweg \etal,
\newblock Phys.\ Lett.{} B~472~(2000)~175\relax
\relax
\bibitem{pl:b511:19}
ZEUS \coll, S.~Chekanov \etal,
\newblock Phys.\ Lett.{} B~511~(2001)~19\relax
\relax
\bibitem{epj:c49:511}
ZEUS \coll, S. Chekanov \etal,
\newblock Eur.\ Phys.\ J.{} C~49~(2007)~511\relax
\relax
\bibitem{epj:c38:437}
H1 \coll, A.~Aktas \etal,
\newblock Eur.\ Phys.\ J.{} C~38~(2004)~437\relax
\relax
\bibitem{epj:c66:17}
H1 \coll, F.D.~Aaron \etal, Eur.\ Phys.\ J.{} C~66~(2010)~17\relax
\bibitem{epj:c54:371}
H1 \coll, F.D.~Aaron \etal,
\newblock Eur.\ Phys.\ J.{} C~54~(2008)~371\relax
\relax
\bibitem{pl:b595:86}
ZEUS \coll, S.~Chekanov \etal,
\newblock Phys.\ Lett.{} B~595~(2004)~86\relax
\relax
\bibitem{pl:b687:16}
ZEUS \coll, S.~Chekanov \etal,
\newblock Phys.\ Lett.{} B~687~(2010)~16\relax
\relax
\bibitem{pl:b715:88}
ZEUS \coll, H. Abramowicz \etal,
\newblock Phys.\ Lett.{} B~715~(2012)~88\relax
\relax
\bibitem{pl:b384:401}
ZEUS Collaboration, M. Derrick \etal,
\newblock Phys.\ Lett.{} B~384 (1996)~401\relax
\relax
\bibitem{epj:c21:303}
M. Fontannaz, J.Ph. Guillet and G. Heinrich, 
\newblock Eur.\ Phys.\ J.{} C~21~(2001)~303\relax
\bibitem{epj:c34:191}
M. Fontannaz and G. Heinrich, 
\newblock Eur.\ Phys.\ J.{} C~34~(2004)~191\relax
\relax
\bibitem{priv:fgh:2013}
M. Fontannaz and G. Heinrich, private communication (2013)\relax
\relax
\bibitem{pr:d72:054002}
A.V.  Lipatov and N.P. Zotov,
\newblock Phys.\ Rev.{} D~72~(2005)~054002\relax
\relax
\bibitem{pr:d81:094027}
A.V.  Lipatov and N.P. Zotov,
\newblock Phys.\ Rev.{} D~81~(2010)~094027\relax
\relax
\bibitem{pr:d88:074001}
A.V. Lipatov, M.A. Malyshev and N.P. Zotov, 
\newblock Phys.\ Rev.\ D~88~(2013) 074001\relax
\relax
\bibitem{lmzpriv}
A.V. Lipatov, M.A. Malyshev and N.P. Zotov, private communication (2014)\relax
\relax
\bibitem{zeus:1993:bluebook}
ZEUS \coll, U.~Holm~(ed.),
\newblock {\em The {ZEUS} Detector}.
\newblock Status Report (unpublished), DESY (1993),
\newblock available on
  \texttt{http://www-zeus.desy.de/bluebook/bluebook.html}\relax
\relax
\bibitem{nim:a279:290}
N.~Harnew \etal,
\newblock Nucl.\ Inst.\ Meth.{} A~279~(1989)~290\relax
\relax
\bibitem{npps:b32:181}
B.~Foster \etal,
\newblock Nucl.\ Phys.\ Proc.\ Suppl.{} B~32~(1993)~181\relax
\relax
\bibitem{nim:a338:254}
B.~Foster \etal,
\newblock Nucl.\ Inst.\ Meth.{} A~338~(1994)~254\relax
\relax
\bibitem{nim:a581:656}
A.~Polini \etal,
\newblock Nucl.\ Inst.\ Meth.{} A~581~(2007)~656\relax
\relax
\bibitem{nim:a309:77}
M.~Derrick \etal,
\newblock Nucl.\ Inst.\ Meth.{} A~309~(1991)~77\relax
\relax
\bibitem{nim:a309:101}
A.~Andresen \etal,
\newblock Nucl.\ Inst.\ Meth.{} A~309~(1991)~101\relax
\relax
\bibitem{nim:a321:356}
A.~Caldwell \etal,
\newblock Nucl.\ Inst.\ Meth.{} A~321~(1992)~356\relax
\relax
\bibitem{nim:a336:23}
A.~Bernstein \etal,
\newblock Nucl.\ Inst.\ Meth.{} A~336~(1993)~23\relax
\relax
\bibitem{lumi2}
L.~Adamczyk \etal, 
\newblock Nucl.\ Inst.\ Meth.{} A~744~(2014)~80\relax
\relax
\bibitem{desy-92-066}
J.~Andruszk\'ow \etal,
\newblock Preprint \mbox{DESY-92-066}, DESY, 1992\relax
\relax
\bibitem{zfp:c63:391}
ZEUS \coll, M.~Derrick \etal,
\newblock Z.\ Phys.{} C~63~(1994)~391\relax
\relax
\bibitem{acpp:b32:2025}
J.~Andruszk\'ow \etal,
\newblock Acta Phys.\ Pol.{} B~32~(2001)~2025\relax
\relax
\bibitem{nim:a565:572}
M.~Helbich \etal,
\newblock Nucl.\ Inst.\ Meth.{} A~565~(2006)~572\relax
\relax
\bibitem{epj:c19:89}
L. Bourhis et al.,
\newblock Eur.\ Phys.\ J.{} C~19~(2001)~89\relax
\relax
\bibitem{jhep:0602:032}
J. Pumplin \etal, \newblock JHEP 02 (2006) 032\relax
\relax
\bibitem{epj:c44:395}
P. Aurenche, M. Fontannaz and J.Ph. Guillet,
\newblock Eur.\ Phys.\ J.{} C~44~(2005) 395\relax
\relax
\bibitem{kmr}
M.A. Kimber, A.D. Martin and M.G. Ryskin,
\newblock Phys.\ Rev.\ D~63~(2001)~114027;\relax
\\
G. Watt, A.D. Martin and M.G. Ryskin,
\newblock Eur.\ Phys.\ J.{} C~31~(2003)~73\relax
\relax
\bibitem{epj:c63:189}
A.D. Martin \etal,
\newblock Eur.\ Phys.\ J.~C~63~(2009) 189\relax
\relax
\bibitem{jhep:0605:026}
T.~Sj\"ostrand \etal,
\newblock JHEP{} 05~(2006)~26\relax
\relax
\bibitem{pr:d55:1280}
H.L. Lai \etal, \relax
\newblock Phys.\ Rev. D 55 (1997) 1280\relax
\relax
\bibitem{grv}
M. Gl\"uck, G. Reya and A. Vogt,
\newblock Phys.\ Rev.\ D~45~(1992)~3986;\relax
\\
M. Gl\"uck, G. Reya and A. Vogt,
\newblock Phys.\ Rev.\ D~46~(1992) 1973\relax
\relax
\bibitem{jhep:0101:010}
G.~Corcella \etal,
\newblock JHEP{} 01~(2001)~010\relax
\relax
\bibitem{tech:cern-dd-ee-84-1}
R.~Brun et al.,
\newblock {\em {\sc geant3}},
\newblock Technical Report CERN-DD/EE/84-1, CERN, 1987\relax
\relax
\bibitem{uproc:chep:1992:222}
W.H.~Smith, K.~Tokushuku and L.W.~Wiggers,
\newblock {\em Proc.\ Computing in High-Energy Physics (CHEP), Annecy, France,
  Sept. 1992}, C.~Verkerk and W.~Wojcik~(eds.), p.~222.
\newblock CERN, Geneva, Switzerland (1992).
\newblock Also in preprint \mbox{DESY 92-150B}\relax
\relax
\bibitem{nim:a580:1257}
P.~Allfrey \etal,
\newblock Nucl.\ Inst.\ Meth.{} A~580~(2007)~1257\relax
\relax
\bibitem{pl:b303:183}
ZEUS \coll, M.~Derrick \etal,
\newblock Phys.\ Lett.{} B~303~(1993)~183\relax
\relax
\bibitem{epj:c1:81}
ZEUS \coll, J.~Breitweg \etal,
\newblock Eur.\ Phys.\ J.{} C~1~(1998)~81\relax
\relax
\bibitem{epj:c6:43}
ZEUS \coll, J.~Breitweg \etal,
\newblock Eur.\ Phys.\ J.{} C~6~(1999)~43\relax
\relax
\bibitem{np:b406:187}
S.~Catani \etal,
\newblock Nucl.\ Phys.{} B~406~(1993)~187\relax
\relax
\bibitem{pr:d48:3160}
S.D.~Ellis and D.E.~Soper,
\newblock Phys.\ Rev.{} D~48~(1993)~3160\relax
\relax
\bibitem{mw}
ZEUS \coll, S. Chekanov \etal,
\newblock  Nucl.\ Phys.{} B~729~(2005)~492;\relax
\\
ZEUS \coll, S. Chekanov \etal, 
\newblock Phys.\ Rev.{} D~76~(2007)~072011;\relax
\\
ATLAS \coll, G. Aad \etal, 
\newblock Nucl.\ Phys.{}  B~875~(2013)~483\relax
\relax
\end{mcbibliography}
\end{document}